\documentclass[american]{iopart}
\usepackage[T1]{fontenc}
\usepackage[utf8]{inputenc}
\usepackage{babel}
\usepackage{prettyref}
\usepackage{bm}
\usepackage{amsbsy}
\usepackage{amstext}
\usepackage{graphicx}
\usepackage[unicode=true,pdfusetitle,
 bookmarks=true,bookmarksnumbered=false,bookmarksopen=false,
 breaklinks=false,pdfborder={0 0 1},backref=false,colorlinks=false]
 {hyperref}
\hypersetup{
 colorlinks=true,linkcolor=blue,citecolor=blue,urlcolor=blue,filecolor=blue}

\makeatletter
\usepackage{iopams}
\usepackage{setstack}

\usepackage[numbers, sort&compress]{natbib}
\usepackage{iopams}

\usepackage{bbold}
\usepackage{stackrel}

\newrefformat{eq}{\hyperref[#1]{Eq.~(\ref{#1})}}
\newrefformat{eqs}{\hyperref[#1]{Eqs.~(\ref{#1})}}
\newrefformat{fig}{\hyperref[#1]{Figure~\ref{#1}}}
\newrefformat{tab}{\hyperref[#1]{Table ~\ref{#1}}}
\newrefformat{sec}{\hyperref[#1]{Section~\ref{#1}}}
\newrefformat{subsec}{\hyperref[#1]{Subsection~\ref{#1}}}
\newrefformat{app}{\hyperref[#1]{Appendix~\ref{#1}}}

\makeatletter
\newsavebox{\@brx}
\newcommand{\llangle}[1][]{\savebox{\@brx}{\(\m@th{#1\langle}\)}%
  \mathopen{\copy\@brx\kern-0.5\wd\@brx\usebox{\@brx}}}
\newcommand{\rrangle}[1][]{\savebox{\@brx}{\(\m@th{#1\rangle}\)}%
  \mathclose{\copy\@brx\kern-0.5\wd\@brx\usebox{\@brx}}}
\makeatother

\newcommand{\newblock}{}

\newcommand{\eqref}{\eref}

\makeatother

\begin{document}
\global\long\def\N{\mathcal{N}}%
\global\long\def\T{\text{T}}%
\global\long\def\D{\mathcal{D}}%
\global\long\def\th{\tilde{h}}%
\global\long\def\tj{\tilde{j}}%
\global\long\def\tx{\tilde{x}}%
\global\long\def\bE{\mathbb{E}}%
\global\long\def\bg{\bm{g}}%
\global\long\def\bh{\bm{h}}%
\global\long\def\bx{\bm{x}}%
\global\long\def\bW{\bm{W}}%
\global\long\def\bR{\mathbb{R}}%
\global\long\def\bj{\bm{j}}%
\global\long\def\tbh{\tilde{\bm{h}}}%
\global\long\def\tbg{\tilde{\bm{g}}}%
\global\long\def\tQ{\tilde{Q}}%
\global\long\def\RNN{\text{RNN}}%
\global\long\def\DNN{\text{DNN}}%

\title{Unified field theoretical approach to deep and recurrent neuronal
networks}
\author{Kai Segadlo$^{1,2,*}$, Bastian Epping$^{2,3,*}$, Alexander van Meegen$^{2,4,*}$,
David Dahmen$^{2}$, Michael Krämer$^{3}$ and Moritz Helias$^{1,2}$}
\address{$^{1}$ Department of Physics, Faculty 1, RWTH Aachen University,
Aachen, Germany}
\address{$^{2}$ Institute of Neuroscience and Medicine (INM-6) and Institute
for Advanced Simulation (IAS-6) and JARA-Institute Brain Structure-Function
Relationships (INM-10), Jülich Research Centre, Jülich, Germany}
\address{$^{3}$ Institute for Theoretical Particle Physics and Cosmology,
RWTH Aachen University, Aachen, Germany}
\address{$^{4}$ Institute of Zoology, University of Cologne, Cologne, Germany}
\address{$^{*}$ Equal contribution}
\begin{abstract}
Understanding capabilities and limitations of different network architectures
is of fundamental importance to machine learning. Bayesian inference
on Gaussian processes has proven to be a viable approach for studying
recurrent and deep networks in the limit of infinite layer width,
$n\to\infty$. Here we present a unified and systematic derivation
of the mean-field theory for both architectures that starts from first
principles by employing established methods from statistical physics
of disordered systems. The theory elucidates that while the mean-field
equations are different with regard to their temporal structure, they
yet yield identical Gaussian kernels when readouts are taken at a
single time point or layer, respectively. Bayesian inference applied
to classification then predicts identical performance and capabilities
for the two architectures. Numerically, we find that convergence towards
the mean-field theory is typically slower for recurrent networks than
for deep networks and the convergence speed depends non-trivially
on the parameters of the weight prior as well as the depth or number
of time steps, respectively. Our method exposes that Gaussian processes
are but the lowest order of a systematic expansion in $1/n$ and we
compute next-to-leading-order corrections which turn out to be architecture-specific.
The formalism thus paves the way to investigate the fundamental differences
between recurrent and deep architectures at finite widths $n$.
\end{abstract}
\maketitle
\tableofcontents{}

\section{Introduction\label{sec:Introduction}}

Deep learning has brought a dramatic improvement of the state-of-the-art
in many fields of data science, ranging from speech recognition and
translation to visual object classification \cite{Hinton06_1527,Krizhevsky12_1097,Hannun14,Lecun2015deep}.
Any progress in the empirically-driven improvement of algorithms must
be accompanied by a profound understanding of why and how deep learning
works. Such an understanding is needed to provide guarantees, for
example about the accuracy and the robustness of the networks, and
will help preventing the frequently reported failures of deep learning,
such as its vulnerability to adversarial examples \cite{Szegedy14_ICLR}.

A common method to obtain analytical insight into deep networks is
to study the overparametrized limit in which the width $n_{\ell}$
of all layers $\ell$ tends to infinity. In this limit, it has been
shown with the help of the multivariate central limit theorem that
under a Gaussian prior on the weights $W^{(\ell)}$ in each layer,
the pre-activations  follow a Gaussian process with an iteratively
determined covariance \cite{Neal96,Williams96_ae5e3ce4,Lee18,Matthews18};
in particular, the pre-activations across different layers and across
different neurons become independently Gaussian distributed. This
approach allows one to investigate learning and prediction in the
framework of Bayesian inference \cite{Williams96_ae5e3ce4}.  There
is a prominent orthogonal line of work based on the neural tangent
kernel \cite{Jacot18_8580} which investigates gradient-based learning
in the large $n_{\ell}$ limit; in this manuscript, we exclusively
focus on Bayesian inference (in this context often called NNGP; see
\cite{Lee20_ad086f59} for an empirical performance comparison of
the two approaches).

Often, analogies are drawn between deep neural networks (DNNs) and
discrete-time recurrent neural networks (RNNs): Unrolling time in
RNNs formally converts them to DNNs, however with shared weights $W^{(\ell)}\equiv W\ \forall\,\ell$
across layers of identical size $n_{\ell}\equiv n\ \forall\,\ell$.
This led to parallel developments in terms of training strategies
for both architectures, such as backpropagation \cite{Rumelhart86_533}
and backpropagation through time \cite{Pearlmutter89}.

There are, however, a number of open issues when applying mean-field
theory to deep and recurrent neural networks. First of all, the approximation
as a Gaussian process relies on the central limit theorem and is thus
strictly valid only in the limit of infinite layer widths $n_{\ell}\to\infty$.
Moreover, due to weight sharing, pre-activations for different points
in time are not statistically independent in RNNs; the central limit
theorem is thus not applicable and the mean-field approximation becomes
uncontrolled. Several studies still find that the mean-field theories
of DNNs and RNNs appear to be closely related, culminating in ref.~\cite{Yang19}
which formulates a variety of network architectures as tensor programs
and finds that most common network architectures, under certain conditions
on the nonlinearities and priors, converge in distribution to a Gaussian
process (see also \cite{Alemohammad21_ICLR,Alemohammad20_04859}
for an extension of the neural tangent kernel to recurrent networks).
But the relationship between the Gaussian processes for RNNs and DNNs
has so far not been addressed. 

The agreement of the mean-field predictions with the performance
of finite-size networks is based on numerical evidence so far. Furthermore,
in the limit of infinite width the number of trainable parameters
of a DNN, $\sum_{\ell=1}^{L}n_{\ell+1}n_{\ell}\to\infty$, and of
an RNN, $n^{2}\to\infty$, both tend to infinity and do not enter
explicitly in the result of the Gaussian approximation. The Gaussian
process thus has limited capability of quantifying the expressivity
of neural networks in relation to the required resources, such as
the number of trained weights. Studies on finite-size corrections
beyond the $n_{\ell}\to\infty$ limit are so far restricted to DNNs
\cite{Yaida20,Dyer20_ICLR,Antognini19_arxiv,Huang20_4542,Aitken20_06687,Halverson21_035002,Naveh21_064301,ZavatoneVeth21_NeurIPS_II,Naveh21_NeurIPS,ZavatoneVeth21_NeurIPS_I,Noci21_DTA7Bgrai-Q,Roberts22}
(but see \cite{Grosvenor22_81} for stationary continuous-time recurrent
networks). Understanding the limits of the putative equivalence of
DNNs and RNNs on the mean-field level requires a common theoretical
basis for the two architectures that would extend to finite $n$ and
finite $n_{\ell}$.

To overcome these limitations, we here combine the established view
of Bayesian inference by Gaussian processes \cite{Williams06} with
statistical field theory applied to neural networks \cite{Sompolinsky88_259,Chow15,Hertz16_033001,Marti18_062314,Crisanti18_062120,Schuecker18_041029}.
These methods have been developed in the field of disordered systems,
which are systems with random parameters, such as spin glasses \cite{Parisi80_1101,Sommers87_1268,Fischer91}
and are able to extract the typical behavior of a system with a large
number of interacting components. For example, this approach has recently
been used to characterize the typical richness, represented by the
entropy, of Boolean functions computed in the output layer of DNNs,
RNNs, and sparse Boolean circuits \cite{Mozeika20_168301}.

Concretely, in this paper, we present a systematic derivation of the
mean-field theories for DNNs and RNNs that is based on the well-established
approach of field theory for recurrent networks \cite{Sompolinsky88_259,molgedey92_3717,Schuecker16b_arxiv,Crisanti18_062120},
which allows a unified treatment of the two architectures \cite{Mozeika20_168301}.
This paves the way for extensions to finite $n,n_{\ell}$, enabled
by a rich set of systematic methods available in the mathematical
physics literature to compute corrections beyond the leading order
\cite{ZinnJustin96,Moshe03}. Already to leading order, we find that
the mean-field theories of DNNs and RNNs are in fact qualitatively
different with regard to correlations across layers or time, respectively.
The predictive distribution in Bayesian training is therefore in general
different between the two architectures. Nonetheless, the structure
of the mean-field equations can give rise to the same Gaussian processes
kernel in the limit of infinite width for both DNNs and RNNs if the
readout in the RNN is taken from a single time step. This finding
holds for single inputs, as pointed out in ref.~\cite{Mozeika20_168301},
as well as input sequences. Furthermore, for a point-symmetric activation
function \cite{Mozeika20_168301}, there is no observable difference
between DNNs and RNNs on the mean-field level if the biases are uncorrelated
in time and the input is only supplied in the first time step. Going
beyond the leading order, we compute the next-to-leading-order corrections
for both DNNs and RNNs. These corrections reveal commonalities and
differences between the architectures: In both architectures external
fluctuations from the input are transmitted by the same linear response
function to the readout layer or time-point. This response decays
exponentially, with identical decay constant with depth or time, respectively.
At the transition to chaos this decay constant diverges. The intrinsically
generated finite-size fluctuations, however, differ with respect to
their layer wise or temporal statistics, respectively. In particular,
cross-time fluctuations drive fluctuations in RNNs overall to a higher
level than in DNNs, where only equal-layer fluctuations drive fluctuations.

\section{Theoretical background}

\subsection{Bayesian supervised learning}

First, we briefly review the Bayesian approach to supervised learning
\cite{Mackay2003}. Let $p(\bm{y}\,|\,\bm{x},\bm{\theta})$ be a model
(here DNN or RNN) that maps inputs $\mathbf{x}\in\mathbb{R}^{n_{\mathrm{in}}}$
to outputs $\bm{y}\in\mathbb{R}^{n_{\text{out}}}$ and that depends
on a set of parameters $\bm{\theta}$. Conventional training of such
a model corresponds to finding a particular parameter set $\hat{\bm{\theta}}$
that maximizes the likelihood $p(\bm{Y}\,|\,\bm{X},\bm{\theta})$
for some given training data $\bm{D}=\{\bm{X},\bm{Y}\}$, with $\bm{X}\in\mathbb{R}^{n_{D}\times n_{\mathrm{in}}}$
and $\bm{Y}\in\bR^{n_{D}\times n_{\mathrm{out}}}$. A prediction for
the output $\bm{y}^{*}$ caused by an unseen test input $\bm{x}^{*}$
is then given by $p(\bm{y}^{*}\,|\,\bm{x}^{*},\hat{\bm{\theta}})$.
In the Bayesian view, one instead assumes a prior distribution of
parameters $p(\bm{\theta})$ to obtain, via Bayes' rule, an entire
posterior distribution of the parameters
\begin{eqnarray}
p(\bm{\theta}\,|\,\bm{Y},\bm{X}) & = & \frac{p(\bm{Y}\,|\,\bm{X},\bm{\theta})\,p(\bm{\theta})}{\int d\bm{\theta}\,p(\bm{Y}\,|\,\bm{X},\bm{\theta})\,p(\bm{\theta})}\,.\label{eq:param_posterior}
\end{eqnarray}
The conditioning on the training data in $p(\bm{\theta}\,|\,\bm{Y},\bm{X})$
can be interpreted as selecting, among all possible parameter sets
given by the prior $p(\bm{\theta})$, those parameter sets that accomplish
the mapping $\bm{X}\rightarrow\bm{Y}$. A Bayesian prediction for
some unseen test input $\bm{x}^{*}$ correspondingly results from
marginalizing the likelihood over the posterior distribution of the
parameters
\begin{eqnarray}
p(\bm{y}^{*}\,|\,\bm{x}^{*},\bm{Y},\bm{X}) & = & \int d\bm{\theta}\,p(\bm{y}^{*}\,|\,\bm{x}^{*},\bm{\theta})\,p(\bm{\theta}\,|\,\bm{Y},\bm{X})\,.\label{eq:param_prediction}
\end{eqnarray}
 Inserting \prettyref{eq:param_posterior} into \prettyref{eq:param_prediction}
yields the predictive distribution 
\begin{eqnarray}
p(\bm{y}^{*}\,|\,\bm{x}^{*},\bm{Y},\bm{X}) & = & \frac{p(\bm{y}^{*},\bm{Y}\,|\,\bm{x}^{*},\bm{X})}{p(\bm{Y}\,|\,\bm{X})}\label{eq:output}
\end{eqnarray}
that depends on the model-dependent \emph{network priors}
\begin{eqnarray}
p(\bm{Y}\,|\,\bm{X}) & = & \int d\bm{\theta}\,p(\bm{Y}\,|\,\bm{X},\bm{\theta})\,p(\bm{\theta}),\label{eq:network_prior}\\
p(\bm{y}^{*},\bm{Y}\,|\,\bm{x}^{*},\bm{X}) & = & \int d\bm{\theta}\,p(\bm{y}^{*}\,|\,\bm{x}^{*},\bm{\theta})\,p(\bm{Y}\,|\,\bm{X},\bm{\theta})\,p(\bm{\theta})\,.\label{eq:network_prior_with_test}
\end{eqnarray}
The network priors encompass all input-output relationships which
are compatible with the prior $p(\bm{\theta})$ and the model. The
difference between the two network priors, \prettyref{eq:network_prior}
and \prettyref{eq:network_prior_with_test}, is the information on
the additional test input $\bm{x}^{*}$ and output $\bm{y}^{*}$.

Note that the Bayesian approach to supervised learning can also be
used for input sequences $\{\mathbf{x}^{(0)},\dots,\mathbf{x}^{(A)}\}$
with $\mathbf{x}^{(a)}\in\mathbb{R}^{n_{\mathrm{in}}}$. To this end,
it is sufficient to replace $\bm{x}\to\{\mathbf{x}^{(0)},\dots,\mathbf{x}^{(A)}\}$
and $\bm{X}\to\{\mathbf{X}^{(0)},\dots,\mathbf{X}^{(A)}\}$ in the
above formulas.

The main difference between Bayesian inference and conventional,
gradient-based training is that in the former one considers an entire
ensemble of networks defined by the posterior distribution of the
parameters while in the latter one seeks a point estimate $\hat{\bm{\theta}}$
for the parameters. Moreover, in the latter approach, learning refers
to the process of refining initial single parameter estimates $\hat{\bm{\theta}}_{0}\to\hat{\bm{\theta}}$
for the approximate posterior $p(\bm{\theta}\,|\,\bm{Y},\bm{X})\approx\delta(\bm{\theta}-\hat{\bm{\theta}}(\bm{Y},\bm{X},\mathcal{L}))$
by minimizing a loss $\mathcal{L}$ on the set of training data $\bm{Y},\bm{X}$.
From the point of view of Bayesian inference, for a posterior that
is sharply peaked at parameter value $\hat{\bm{\theta}}$, a Laplace
approximation leads to a prediction based on the maximum a posteriori
estimate, $p(\bm{y}^{*}\,|\,\bm{x}^{*},\bm{Y},\bm{X})\approx p(\bm{y}^{*}\,|\,\bm{x}^{*},\hat{\bm{\theta}})$.
Thus, if $\hat{\bm{\theta}}$ corresponds to the parameter value found
by conventional training, the Bayesian prediction and the prediction
based on conventional training coincide.

In the following, we use a field theoretic approach to calculate the
network priors for both deep and recurrent neural networks. Conditioning
on the training data, \prettyref{eq:output}, then yields the Bayesian
prediction of the output.

\subsection{Network architectures\label{subsec:Network-architectures}}

Deep feedforward neural networks (DNNs) and discrete-time recurrent
neural networks (RNNs) can both be described by a set of pre-activations
$\bm{h}^{(a)}\in\bR^{n_{a}}$ that are determined by an affine linear
transformation
\begin{eqnarray}
\bm{h}^{(a)} & = & \bm{W}^{(a)}\phi(\bm{h}^{(a-1)})+\bm{W}^{(\text{in},a)}\bm{x}^{(a)}+\bm{\xi}^{(a)}\label{eq:eq_of_motion}
\end{eqnarray}
of activations $\phi(\bm{h}^{(a-1)})\in\bR^{n_{a-1}}$. The pre-activations
are transformed by an activation function $\phi:\bR\rightarrow\bR$
which is applied element-wise to vectors. For DNNs, $\bm{W}^{(a)}\in\bR^{n_{a}\times n_{a-1}}$
denotes the weight matrix from layer $a-1$ to layer $a$, and $\bm{\xi}^{(a)}\in\bR^{n_{a}}$
represents biases in layer $a$. Inputs $\bm{x}^{(a)}$ are typically
only applied to the first layer such that the input matrices $\bm{W}^{(\text{in},a)}\in\bR^{n_{a}\times n_{\mathrm{in}}}$
vanish for $a>0$. For RNNs, the index $a$ denotes different time
steps. The weight matrix, input matrix, and biases are static over
time, $\bm{W}^{(a)}\equiv\bm{W}$, $\bm{W}^{(\text{in},a)}\equiv\bm{W}^{(\text{in})}$,
and $\bm{\xi}^{(a)}\equiv\bm{\xi}$, and couple activities across
successive time steps. For both architectures, we include an additional
input and output layer 
\begin{eqnarray}
\bm{h}^{(0)} & = & \bm{W}^{(\mathrm{in},0)}\bm{x}^{(0)}+\bm{\xi}^{(0)},\label{eq:input_layer}\\
\bm{y} & = & \bm{W}^{(\mathrm{out})}\phi(\bm{h}^{(A)})+\bm{\xi}^{(A+1)},\label{eq:output_layer}
\end{eqnarray}
with $\bm{W}^{(\mathrm{out})}\in\bR^{n_{\mathrm{out}}\times n_{A}}$,
which allow us to set independent input and output dimensions. Here,
$A$ denotes the final layer for the DNN and the final time point
for the RNN. The set of parameters $\bm{\theta}$ is the collection
of $\bm{W}^{(\mathrm{in},a)},\bm{W}^{(\mathrm{out})},\bm{W}^{(a)},$
and $\bm{\xi}^{(a)}$.

\subsection{Parameter priors}

We use Gaussian priors for all model parameters, that is for the RNN
\begin{eqnarray}
W_{ij} & \stackrel{\text{i.i.d.}}{\sim} & \N(0,n^{-1}g^{2}),\\
W_{ij}^{(\mathrm{in})} & \stackrel{\text{i.i.d.}}{\sim} & \N(0,n_{\mathrm{in}}^{-1}g_{0}^{2}),\\
\xi_{i} & \stackrel{\text{i.i.d.}}{\sim} & \N(0,\sigma^{2}),
\end{eqnarray}
and for the DNN
\begin{eqnarray}
W_{ij}^{(a)} & \stackrel{\text{i.i.d.}}{\sim} & \N(0,n_{a-1}^{-1}g_{a}^{2}),\\
W_{ij}^{(\mathrm{in},a)} & \stackrel{\text{i.i.d.}}{\sim} & \N(0,n_{\mathrm{in}}^{-1}g_{0}^{2}),\\
\xi_{i}^{(a)} & \stackrel{\text{i.i.d.}}{\sim} & \N(0,\sigma_{a}^{2}),
\end{eqnarray}
as well as 
\begin{eqnarray}
W_{ij}^{(\mathrm{out})} & \stackrel{\text{i.i.d.}}{\sim} & \N(0,n_{\mathrm{A}}^{-1}g_{A+1}^{2}),\\
\xi_{i}^{(A+1)} & \stackrel{\text{i.i.d.}}{\sim} & \N(0,\sigma_{A+1}^{2}),
\end{eqnarray}
for both architectures (where $n_{A}=n$ for the RNN). These priors
on the parameters are used to calculate the network prior $p(\bm{Y}\,|\,\bm{X})$.

\section{Unified field theoretical approach to RNNs and DNNs\label{sec:Unified-field-theory}}

The network prior $p(\bm{Y}\,|\,\bm{X})$, \prettyref{eq:network_prior},
is a joint distribution of all outputs $\bm{y}_{\alpha}\in\bm{Y}$,
each corresponding to a single training input $\bm{x}_{\alpha}\in\bm{X}$.
Its calculation is tantamount to a known problem in physics, the replica
calculation \cite{Hertz91,ZinnJustin96}. Here, each replicon is a
copy of the network with the same parameters $\bm{\theta}$ but a
different input $\bm{x}_{\alpha}$. For simplicity, in the following
we illustrate the derivation of $p(\bm{y}\,|\,\bm{x})$ for a single
input $\bm{x}\equiv\bm{x}^{(a=0)}$ that is presented to the first
layer of the DNN or at the first time point for the RNN, respectively.
We present the more cumbersome but conceptually similar general case
of multiple inputs, or multiple input sequences, in \prettyref{app:replicated-Field-theory-DNN-RNN}.

The network prior is defined as the probability of the output given
the input, 
\begin{eqnarray}
p(\bm{y}\,|\,\bm{x}) & = & \int d\bm{\theta}\,p(\bm{y}\,|\,\bm{x},\bm{\theta})\,p(\bm{\theta}),\label{eq:prior_single_input}
\end{eqnarray}
marginalized over the parameter prior, where
\begin{eqnarray}
p(\bm{y}\,|\,\bm{x},\bm{\theta})=\int d\bm{h}^{(0)} & \dots & \int d\bm{h}^{(A)}\,\delta\left(\bm{y}-\bm{W}^{(\text{out})}\bm{\phi}^{(A)}-\bm{\xi}^{(A+1)}\right)\nonumber \\
 & \times & \prod_{a=1}^{A}\delta\left(\bm{h}^{(a)}-\bm{W}^{(a)}\bm{\phi}^{(a-1)}-\bm{\xi}^{(a)}\right)\nonumber \\
 & \times & \delta\left(\bm{h}^{(0)}-\bm{W}^{(\text{in})}\bm{x}-\bm{\xi}^{(0)}\right),\label{eq:Ansatz}
\end{eqnarray}
follows by enforcing the set of equations, \prettyref{eq:eq_of_motion}
to \prettyref{eq:output_layer}, using Dirac constraints. Throughout
the manuscript, we use the abbreviation $\bm{\phi}^{(a)}=\phi(\bm{h}^{(a)})$.

\subsection{Marginalization of the parameter prior}

From \prettyref{eq:Ansatz}, it follows that the computation of the
marginalization of the parameters $\bm{\theta}$ in \prettyref{eq:prior_single_input}
can be reduced to
\begin{eqnarray}
p(\bm{y}\,|\,\bm{x})=\int d\bm{h}^{(0)} & \dots & \int d\bm{h}^{(A)}\nonumber \\
 & \times & \left\langle \delta\left(\bm{y}-\bm{W}^{(\text{out})}\bm{\phi}^{(A)}-\bm{\xi}^{(A+1)}\right)\right\rangle _{\bm{W}^{(\text{out})},\bm{\xi}^{(A+1)}}\nonumber \\
 & \times\Big\langle & \Big\langle\prod_{a=1}^{A}\delta\left(\bm{h}^{(a)}-\bm{W}^{(a)}\bm{\phi}^{(a-1)}-\bm{\xi}^{(a)}\right)\Big\rangle_{\{\bm{W}^{(a)}\}}\nonumber \\
 & \times & \left\langle \delta\left(\bm{h}^{(0)}-\bm{W}^{(\text{in})}\bm{x}-\bm{\xi}^{(0)}\right)\right\rangle _{\bm{W}^{(\text{in})}}\Big\rangle_{\{\bm{\xi}^{(a)}\}}.
\end{eqnarray}
To proceed, it is advantageous to represent the Dirac $\delta$-distributions
as Fourier integrals,
\begin{eqnarray}
\delta(\bm{h}) & = & \int d\tilde{\bm{h}}\,\exp(\tilde{\bm{h}}^{\T}\bm{h})\label{eq:Dirac_Fourier}
\end{eqnarray}
with the inner product $\tilde{\bm{h}}^{\T}\bm{h}=\sum_{k}\tilde{h}_{k}h_{k}$
and $\int d\tilde{\bm{h}}=\prod_{k}\int_{i\mathbb{R}}\frac{d\tilde{h}_{k}}{2\pi i}$,
because it leads to averages of the form $\langle\exp(k\theta)\rangle_{\theta}$
which are analytically solvable. Using $\langle\exp(k\theta)\rangle_{\theta}=\exp\left(\frac{1}{2}\sigma^{2}k^{2}\right)$
for $\theta\sim\mathcal{N}(0,\sigma^{2})$, the network prior for
a single replicon, $p(\bm{y}\,|\,\bm{x})$, takes the form (details
in \prettyref{app:replicated-Field-theory-DNN-RNN})
\begin{eqnarray}
p(\bm{y}\,|\,\bm{x}) & = & \int d\tilde{\bm{y}}\int\D\bm{h}\int\D\tilde{\bm{h}}\,\exp\left(\mathcal{S}(\bm{y},\tilde{\bm{y}},\bm{h},\tilde{\bm{h}}\,|\,\bm{x})\right),\label{eq:moment_gen_function}
\end{eqnarray}
where $\int\D\bm{h}\equiv\prod_{a=0}^{A}\prod_{k}\int_{\mathbb{R}}dh_{k}^{(a)}$
and $\int\D\tilde{\bm{h}}\equiv\prod_{a=0}^{A}\prod_{k}\int_{i\mathbb{R}}\frac{d\tilde{h}_{k}^{(a)}}{2\pi i}$.
The exponent $\mathcal{S}$, commonly called \emph{the action,} is
given by
\begin{eqnarray}
\mathcal{S}(\bm{y},\tilde{\bm{y}},\bm{h},\tilde{\bm{h}}\,|\,\bm{x}) & = & \mathcal{S}_{\text{out}}(\bm{y},\tilde{\bm{y}}\,|\,\bm{h}^{(A)})+\mathcal{S}_{\text{net}}(\bm{h},\tilde{\bm{h}}\,|\,\bm{x}),
\end{eqnarray}
where
\begin{eqnarray}
\mathcal{S}_{\text{net}}(\bm{h},\tilde{\bm{h}}\,|\,\bm{x}) & := & \sum_{a=0}^{A}\tilde{\bm{h}}^{(a)\T}\bm{h}^{(a)}+\frac{1}{2}\sum_{a,b=0}^{A}\sigma_{a}^{2}\,\tilde{\bm{h}}^{(a)\T}M_{a,b}\tilde{\bm{h}}^{(b)}\nonumber \\
 &  & +\frac{1}{2}\sum_{a,b=1}^{A}\frac{g_{a}^{2}}{n_{a-1}}\tilde{\bm{h}}^{(a)\T}\bm{\phi}^{(a-1)\T}M_{a,b}\bm{\phi}^{(b-1)}\tilde{\bm{h}}^{(b)}\nonumber \\
 &  & +\frac{g_{0}^{2}}{2n_{\text{in}}}\tilde{\bm{h}}^{(0)\T}\bm{\bm{x}}^{\T}\bm{\bm{x}}\tilde{\bm{h}}^{(0)}\label{eq:S_rec}
\end{eqnarray}
is the action of the input and the recurrent layer of the RNN or the
inner part of the DNN, respectively, and 
\begin{eqnarray}
\mathcal{S}_{\text{out}}(\bm{y},\tilde{\bm{y}}\,|\,\bm{h}^{(A)}) & := & \tilde{\bm{y}}^{\T}\bm{y}+\frac{\sigma_{A+1}^{2}}{2}\tilde{\bm{y}}^{\T}\tilde{\bm{y}}+\frac{g_{A+1}^{2}}{2n_{A}}\tilde{\bm{y}}^{\T}\bm{\phi}^{(A)\T}\bm{\phi}^{(A)}\tilde{\bm{y}}
\end{eqnarray}
is the action for the output layer. Note that $\mathcal{S}$ is diagonal
in neuron indices with respect to the explicitly appearing fields
$\bm{h}$ and $\tilde{\bm{h}}$ and couplings across neurons are only
mediated by terms of the form $\propto\bm{\phi}^{\T}\bm{\phi}$.

For RNNs, the shared connectivity and biases at different time points
imply correlations across time steps; for DNNs, in contrast, the connectivity
and biases are realized independently across layers, so that the action
decomposes into a sum of $A+2$ individual layers. In \prettyref{eq:S_rec},
this leads to

\begin{eqnarray} M_{a,b}& =& \cases{1&\RNN \\ \delta_{a,b} & \DNN} \label{eq:net_distinguishing_matrix} \end{eqnarray}which
is the only difference between DNN and RNN in this formalism.

\subsection{Auxiliary variables}

An action that is quadratic in $\bm{h}$ and $\tilde{\bm{h}}$ corresponds
to a Gaussian and therefore to an analytically solvable integral.
However, the post-activations $\bm{\phi}\equiv\phi(\bm{h})$ in $\mathcal{S}_{\mathrm{net}}$
and $\mathcal{S}_{\text{out}}$ introduce a non-quadratic part and
the terms $\propto\,\tilde{\bm{h}}^{\T}\tilde{\bm{h}}\,\bm{\phi}^{\T}\bm{\phi}$
cause a coupling across neurons. To deal with this difficulty, we
introduce new auxiliary variables 
\begin{eqnarray}
C^{(a,b)} & := & M_{a,b}\left[\sigma_{a}^{2}+\mathbb{1}_{a\ge1,b\ge1}\frac{g_{a}^{2}}{n_{a-1}}\bm{\phi}^{(a-1)\T}\bm{\phi}^{(b-1)}\right]\nonumber \\
 &  & +\mathbb{1}_{a=0,b=0}\frac{g_{0}^{2}}{n_{\text{in}}}\bm{x}^{\T}\bm{x},\label{eq:def_aux_field}
\end{eqnarray}
where $0\le a,b\le A+1$, a common practice originating from dynamic
spin-glass theory \cite{Sompolinsky81} and used for random networks
\cite{Crisanti18_062120,Marti18_062314,Schuecker18_041029,Helias20_970}.
The second term $\propto\bm{\phi}^{\T}\bm{\phi}$ in \prettyref{eq:def_aux_field}
contains the sum of post-activations over all neuron indices, where
the indicator function $\mathbb{1}$ is $1$ if both conditions in
the subscripts are fulfilled and $0$ otherwise. Assuming sufficiently
weak correlations among the $\phi_{i}$, we expect the sum to be close
to its mean value with decreasing variations as $n_{a}$ grows; for
large $n_{a}$ the sum is thus close to a Gaussian. This intuition
is made precise below by a formal saddle point approximation in $C$.

Enforcing the auxiliary variables through Dirac-$\delta$ constraints,
analogous to \prettyref{eq:Dirac_Fourier} (see \prettyref{app:replicated-Field-theory-DNN-RNN}
for details), leads to the prior distribution

\begin{eqnarray}
p(\bm{y}\,|\,\bm{x}) & = & \int d\tilde{\bm{y}}\,\exp\left(\tilde{\bm{y}}^{\T}\bm{y}\right)\left\langle \exp\left(\frac{1}{2}\tilde{\bm{y}}^{\T}C^{(A+1,A+1)}\tilde{\bm{y}}\right)\right\rangle _{C,\tilde{C}},\label{eq:single_output_prob}
\end{eqnarray}
where the distribution $C,\tilde{C}\sim\exp\left(\mathcal{S}_{\mathrm{aux}}(C,\tilde{C})\right)$
is described by the action

\begin{eqnarray}
\mathcal{S}_{\mathrm{aux}}(C,\tilde{C}) & := & -n\sum_{a,b=0}^{A+1}\nu_{a-1}\,\tilde{C}^{(a,b)}C^{(a,b)}+n\,\mathcal{W}_{\mathrm{aux}}(\tilde{C}\,|\,C).\label{eq:RNN_decoupled_action}
\end{eqnarray}
To account for possible differences in layer widths in DNNs, we here
introduce $\nu_{a}=n_{a}/n=\mathcal{O}(1)$ as the relative network
widths with respect to some reference $n$. For RNNs, all layers have
the same width so that $\nu_{a>0}=1$. For the input we have $n_{-1}\equiv n_{\text{in}}$.
Lastly, we assume that the output size $n_{\mathrm{out}}$ does not
scale with $n$, i.e., $n_{\mathrm{out}}=O(1)$, such that the only
$O(n)$ contribution in the $\langle\cdot\rangle_{C,\tilde{C}}$ expectation
in \prettyref{eq:single_output_prob} is due to \prettyref{eq:RNN_decoupled_action}.
The recurrent part and the input, which decouple in the neurons, are
together described for the DNN by

\begin{eqnarray}
\mathcal{W}_{\text{aux}}^{\DNN}(\tilde{C}\,|\,C) & = & \sum_{a=1}^{A+1}\nu_{a-1}\ln\left\langle e^{\tilde{C}^{(a,a)}\,g_{a}^{2}\phi^{(a-1)}\phi^{(a-1)}}\right\rangle _{h^{(a-1)}}\nonumber \\
 &  & +\nu_{-1}\,\tilde{C}^{(0,0)}\frac{g_{0}^{2}}{n_{\text{in}}}\bm{x}^{\T}\bm{x}+\sum_{a=0}^{A+1}\nu_{a-1}\,\tilde{C}^{(a,a)}\sigma_{a}^{2}\label{eq:W_rec_DNN}
\end{eqnarray}
with $h^{(a)}\sim\N(0,C^{(a,a)})$ a scalar centered Gaussian with
layer-dependent variance $\langle h^{(a)}h^{(a)}\rangle=C^{(a,a)}$
and for the RNN by
\begin{eqnarray}
\mathcal{W}_{\text{aux}}^{\RNN}(\tilde{C}\,|\,C) & = & \ln\left\langle e^{\sum_{a,b=1}^{A+1}\tilde{C}^{(a,b)}\,g^{2}\phi^{(a-1)}\phi^{(b-1)}}\right\rangle _{\{h^{(a)}\}}\nonumber \\
 &  & +\nu_{-1}\,\tilde{C}^{(0,0)}\frac{g_{0}^{2}}{n_{\text{in}}}\bm{x}^{\T}\bm{x}+\sum_{a,b=0}^{A+1}\tilde{C}^{(a,b)}\sigma^{2}\label{eq:W_rec_RNN}
\end{eqnarray}
with $\{h^{(a)}\}\equiv\{h^{(a)}\}_{0\le a\le A}$ and $\{h^{(a)}\}_{0\le a\le A}\sim\N(0,C)$
a scalar centered Gaussian across time with covariance matrix $\langle h^{(a)}h^{(b)}\rangle=C^{(a,b)}$.
Note that the $\phi$ in \eqref{eq:W_rec_DNN} and \eqref{eq:W_rec_RNN}
are one-dimensional with regard to the neuron index, since the system
is homogeneous across neurons after the disorder average; the number
of neurons is contained in the prefactor $n$ in \eqref{eq:RNN_decoupled_action}
and in the factors $\nu_{a}$ in \eqref{eq:W_rec_DNN}.

\subsection{Saddle-point approximation}

The factor $n$ in \prettyref{eq:RNN_decoupled_action}, which stems
from the decoupling across neurons, for large $n$ leads to a strongly
peaked distribution of $C$ and $\tilde{C}$. Therefore we can use
a saddle point approximation to calculate the average over $C$ and
$\tilde{C}$ in \prettyref{eq:single_output_prob}. In the limit $n\to\infty$
this approximation becomes exact.

We thus search for stationary points of the action
\begin{eqnarray}
\frac{\partial}{\partial C^{(a,b)}}\mathcal{S}_{\mathrm{aux}}(C,\tilde{C}) & \stackrel{!}{=} & 0,\label{eq:saddle_action}\\
\frac{\partial}{\partial\tilde{C}^{(a,b)}}\mathcal{\mathcal{S}_{\mathrm{aux}}}(C,\tilde{C}) & \stackrel{!}{=} & 0,\nonumber 
\end{eqnarray}
which yields a coupled set of self-consistency equations for the mean
values $\overline{C}$ and $\overline{\tilde{C}}$, commonly called
mean-field equations: $\overline{\tilde{C}}^{(a,b)}\equiv0$, which
follows from the normalization of the probability distribution \cite{Coolen01},
and

\begin{eqnarray}
\overline{C}{}^{(a,b)} & = & M_{a,b}\left[\sigma_{a}^{2}+\mathbb{1}_{a\ge1,b\ge1}\,g_{a}^{2}\langle\phi(h^{(a-1)})\phi(h^{(b-1)})\rangle_{h^{(a-1)},h^{(b-1)}}\right]\nonumber \\
 &  & +\mathbb{1}_{a=0,b=0}\,\frac{g_{0}^{2}}{n_{\text{in}}}\bm{x}^{\T}\bm{x}\label{eq:MFT}
\end{eqnarray}
with $h^{(a-1)},h^{(b-1)}\sim\mathcal{N}(0,\overline{C})$. \prettyref{eq:MFT}
comprises both DNN and RNN; the difference between \prettyref{eq:W_rec_DNN}
and \prettyref{eq:W_rec_RNN} leads to the appearance of $M_{a,b}$
on the r.h.s. The average on the r.h.s.~has to be taken with respect
to a probability distribution that only includes two layers or time
points. This is due to the marginalization property of the Gaussian
distribution of the pre-activations $h^{(a-1)}$, which results from
inserting the saddle-point solutions \prettyref{eq:MFT} for $\overline{C}$
and $\overline{\tilde{C}}.$ Accordingly, we are left with a closed
system of equations for the saddle-point values $\overline{C}$ that
are the layer- or time-dependent correlations. These equations need
to be solved recursively from the input $\overline{C}^{(0,0)}=\sigma_{0}^{2}+\frac{g_{0}^{2}}{n_{\text{in}}}\bm{x}^{\T}\bm{x}$
to the output $\overline{C}^{(A+1,A+1)}=\sigma_{A+1}^{2}+g_{A+1}^{2}\langle\phi(h^{(A)})\phi(h^{(A)})\rangle_{h^{(A)},h^{(A)}}$.

\subsection{Network prior}

Computing the Gaussian integral over $\tilde{\bm{y}}$ in the saddle-point
approximation of \prettyref{eq:single_output_prob}, one obtains the
distribution of the outputs as independent Gaussians across neurons
$i$
\begin{equation}
p(\bm{y}\,|\,\bm{x})=\prod_{i}p(y_{i}\,|\,\bm{x})=\prod_{i}\N(y_{i}\,;0,\overline{C}^{(A+1,A+1)})\,.
\end{equation}
An analogous calculation for multiple input sequences $\{\bm{x}_{\alpha}^{(0)},\dots,\bm{x}_{\alpha}^{(A)}\}$
(see \prettyref{app:replicated-Field-theory-DNN-RNN}) yields the
equivalent mean-field equations

\begin{eqnarray}
\overline{C}_{\alpha\beta}^{(a,b)} & =M_{a,b}\bigg[ & \sigma_{a}^{2}+\mathbb{1}_{a\ge1,b\ge1}\,g_{a}^{2}\langle\phi(h_{\alpha}^{(a-1)})\phi(h_{\beta}^{(b-1)})\rangle_{h_{\alpha}^{(a-1)},h_{\beta}^{(b-1)}}\nonumber \\
 &  & +\mathbb{1}_{a\le A,b\le A}\,\frac{g_{0}^{2}}{n_{\text{in}}}\bm{x}_{\alpha}^{(a)\T}\bm{x}_{\beta}^{(b)}\bigg]\label{eq:MFT_replicated}
\end{eqnarray}
with $h_{\alpha}^{(a-1)},h_{\beta}^{(b-1)}\sim\N(0,\overline{C})$
and $0\le a,b\le A+1$. These lead to the joint network prior
\begin{eqnarray}
p(\bm{Y}\,|\,\{\bm{X}^{(0)},\dots,\bm{X}^{(A)}\}) & = & \prod_{i}p(\boldsymbol{y}_{i}\,|\,\{\bm{X}^{(0)},\dots,\bm{X}^{(A)}\})\nonumber \\
 & = & \prod_{i}\N(\boldsymbol{y}_{i}\,;0,\boldsymbol{K})
\end{eqnarray}
where the covariance matrix is the Gram matrix of the kernel \cite{Williams06},
\begin{eqnarray}
K_{\alpha\beta} & = & \overline{C}{}_{\alpha,\beta}^{(A+1,A+1)}\,.
\end{eqnarray}
Here $\boldsymbol{y}_{i}$ denotes the $i$-th row of the output matrix
$\bm{Y}$ that comprises the output of neuron $i$ to all input sequences
$\{\bm{x}_{\alpha}^{(0)},\dots,\bm{x}_{\alpha}^{(A)}\}$.

In principle, it is also possible to use independent biases or input
weights across time steps in the RNN. This would lead to the respective
replacements $M_{a,b}\sigma^{2}\to\delta_{a,b}\sigma^{2}$ and $M_{a,b}\mathbb{1}_{a\le A,b\le A}\,\frac{g_{0}^{2}}{n_{\text{in}}}\bm{x}_{\alpha}^{(a)\T}\bm{x}_{\beta}^{(b)}\to\delta_{a,b}\mathbb{1}_{a\le A,b\le A}\,\frac{g_{0}^{2}}{n_{\text{in}}}\bm{x}_{\alpha}^{(a)\T}\bm{x}_{\beta}^{(b)}$
in \prettyref{eq:MFT_replicated}.

\subsection{Predictive distribution}

We split $\boldsymbol{X},\bm{Y}$ into training data (indexed by subscript
$D$) and test data (indexed by subscript $*$). The conditioning
on the training data via \prettyref{eq:output} can here be done analytically
because the network priors are Gaussian \cite{Williams06}. For scalar
inputs, this yields the predictive distribution
\begin{equation}
p(\bm{Y}_{*}\,|\,\bm{X}_{*},\bm{Y}_{D},\bm{X}_{D})=\prod_{i}\N(\boldsymbol{y}_{*i}\,;\bm{\mu}_{GP},\bm{K}_{GP})\label{eq:predictive_dist_final}
\end{equation}
with
\begin{eqnarray}
\bm{\mu}_{GP} & = & \bm{K}_{*D}\bm{K}_{DD}^{-1}\bm{y}_{D},\qquad\bm{K}_{GP}=\bm{K}_{**}-\bm{K}_{*D}\bm{K}_{DD}^{-1}\bm{K}_{*D}^{T},\label{eq:GP_prediction-2}
\end{eqnarray}
which are fully determined by the kernel matrix $\bm{K}=\left(\begin{array}{cc}
\bm{K}_{DD} & \bm{K}_{*D}^{T}\\
\bm{K}_{*D} & \bm{K}_{DD}
\end{array}\right)\,$. For input sequences, it is again sufficient to replace $\bm{X}_{*}\to\{\bm{X}_{*}^{(0)},\dots,\bm{X}_{*}^{(A)}\}$
and $\bm{X}_{D}\to\{\bm{X}_{D}^{(0)},\dots,\bm{X}_{D}^{(A)}\}$.

\section{Comparison of RNNs and DNNs\label{sec:Comparison-of-RNN-and-DNN}}

\begin{figure}
\begin{centering}
\includegraphics{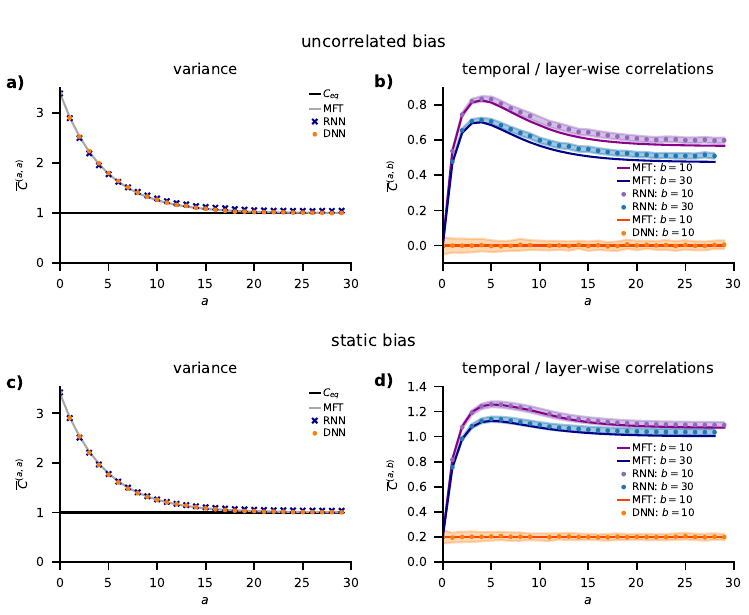}
\par\end{centering}
\caption{\textbf{Mean-field theory for DNN and RNN with a single input. a)
}Average variance in mean-field theory $\overline{C}{}^{(a,a)}$ (\prettyref{eq:MFT};
solid gray curve) and estimate $\frac{1}{n_{a}}\sum_{i}h_{i}^{(a)}h_{i}^{(a)}$
from simulation, averaged over $100$ realizations of networks, for
biases that are uncorrelated across time/layers (blue crosses RNN;
orange dots DNN). \textbf{b) }Cross-covariance $\overline{C}{}^{(a,b)}$
as a function of the hidden layer index $a$ for fixed $b\in\{10,30\}$
and uncorrelated biases. RNN: Mean-field theory (solid dark blue and
dark magenta). Mean (blue / purple dots) and standard error of the
mean (light blue / light purple tube) of $\frac{1}{n_{a}}\sum_{i}h_{i}^{(a)}h_{i}^{(b)}$
estimated from simulation of $100$ network realizations. DNN: Mean
(orange dots) and standard error of the mean of $\frac{1}{n_{a}}\sum_{i}h_{i}^{(a)}h_{i}^{(b)}$
estimated from simulation of $100$ network realizations. Other parameters
$g_{0}^{2}=g^{2}=1.6$, $\sigma^{2}=0.2$, finite layer width $n_{a}=2000$,
$A=30$ hidden layers, ReLU activation $\phi(x)=\max(0,x)$ and Gaussian
inputs $\bm{x}\protect\overset{\text{i.i.d.}}{\sim}\mathcal{N}(1,1)$
with $n_{\text{in}}=10^{5}$. \textbf{c)} Same as a) but for biases
that are static across time/layers. \textbf{d)} Same as b) but for
the static bias case.\label{fig:Mean-field-theory-for-DNN-RNN}}
\end{figure}

Above, we derived the mean-field equations \eqref{eq:MFT_replicated}
for the kernel matrix $\boldsymbol{K}$ using a field-theoretic approach.
Here, we investigate differences in the mean-field distributions of
the different network architectures, starting with the kernel and
considering the predictive distribution afterwards.

\subsection{Kernel\label{sec:Kernel}}

The diagonal elements, $\overline{C}{}^{(a,a)}$ for the single-input
case in \prettyref{eq:MFT} and equivalently $\overline{C}{}_{\alpha,\beta}^{(a,a)}$
for the multiple-input-sequences case in \prettyref{eq:MFT_replicated},
are identical for RNNs and DNNs, because $M_{a,a}=1$ for both architectures.
This implies that the equal-time or within-layer statistics, correspondingly,
is the same in both architectures. The reason is that the iterations
\prettyref{eq:MFT} and \prettyref{eq:MFT_replicated} for equal-time
points $a=b$ form closed sets of equations; they can be solved independently
of the statistics for different time points $a\neq b$. Formally,
this follows from the marginalization property of the Gaussian, which
implies that any subset of a multivariate Gaussian is Gaussian, too,
with a covariance matrix that is the corresponding sector of the covariance
matrix of all variables \cite{Williams06}. The precise agreement
of this mean-field prediction with the average correlation estimated
from direct simulation is shown in \prettyref{fig:Mean-field-theory-for-DNN-RNN}a
and c for the single-input case for both uncorrelated (a) and static
biases (c) across time or layers, respectively.

A notable difference between RNN and DNN is that activity in the RNN
is correlated across time steps due to the shared weights, even if
biases are uncorrelated in time, as shown in \prettyref{fig:Mean-field-theory-for-DNN-RNN}b.
Static biases simply strengthen the correlations across time steps
(see \prettyref{fig:Mean-field-theory-for-DNN-RNN}d). For DNNs, in
contrast, cross-layer correlations only arise due to biases that are
correlated across layers, because weights are drawn independently
for each layer. This is shown in \prettyref{fig:Mean-field-theory-for-DNN-RNN}b
and d: Correlations vanish for DNNs in the uncorrelated bias case
(b) and take on the value $\sigma^{2}$, the variance of the bias,
in the static bias case (d). Again, the mean-field theory accurately
predicts the non-zero correlations across time in the RNN as well
as the correlations across layers generated by the correlated biases
in the DNN. In the RNN, temporal correlations show a non-trivial interplay
due to the shared weights across time. We observe an instability that
can build up by this mechanism in finite-size RNNs, even in parameter
regimes that are deemed stable in mean-field theory (see \prettyref{app:RNN_instability},
\prettyref{fig:finite_size_effect}).

In a particular case, the correlations across time steps also vanish
for the RNN: we show by induction that off-diagonal elements vanish
for point-symmetric activation functions if inputs are only provided
in the initial time step, $\{\bm{X}^{(0)},0,\dots,0\}\equiv\bm{X}$,
and the bias is absent, $\sigma=0$ (or uncorrelated across time steps).
Assuming that $\overline{C}_{\alpha,\beta}^{(a-1,b-1)}\stackrel{a\neq b}{=}0$,
we have

\begin{eqnarray}
\overline{C}{}_{\alpha,\beta}^{(a,b)} & = & g^{2}\,\langle\phi(h_{\alpha}^{(a-1)})\rangle_{h_{\alpha}^{(a-1)}}\langle\phi(h_{\beta}^{(b-1)})\rangle_{h_{\beta}^{(b-1)}}\overset{\phi\text{ odd}}{=}0\label{eq:C_a_neq_b}
\end{eqnarray}
with $h_{\alpha}^{(a-1)}\sim\N(0,\overline{C})$ and $h_{\beta}^{(b-1)}\sim\N(0,\overline{C})$.
Hence, if the pre-activations $h_{\alpha}^{(a-1)},h_{\beta}^{(b-1)}$
at points $a-1$ and $b-1$ are uncorrelated, also $h_{\alpha}^{(a)},h_{\beta}^{(b)}$
will be uncorrelated. The base case of the induction proof follows
from the independence of the input weights $\bm{W}^{(\mathrm{in})}$
and the recurrent weights $\bm{W}$: Correlations between time point
zero and other time points are zero. Therefore, by induction in time,
time points will be uncorrelated at any point, meaning that for odd
activations $\phi$ and the considered input layer, the solutions
of the mean-field equations are the same for DNNs and RNNs. \prettyref{fig:Mean-field-theory-for-DNN-RNN-erf}
in the \prettyref{app:erf_appendix} is similar to \prettyref{fig:Mean-field-theory-for-DNN-RNN}
but for the $\mathrm{erf}$ nonlinearity. There we observe the vanishing
temporal correlation for RNNs with uncorrelated bias explicitly.

\subsection{Predictive distribution}

\begin{figure}
\begin{centering}
\includegraphics{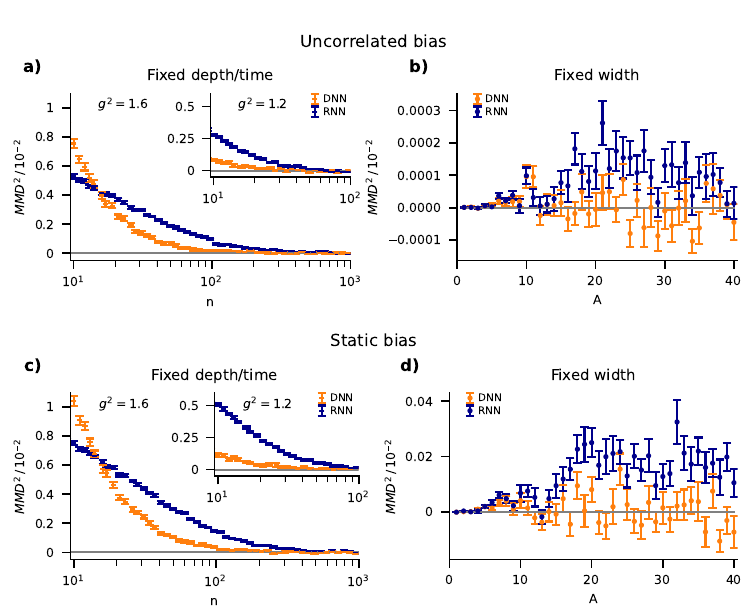}
\par\end{centering}
\caption{\textbf{Convergence of RNN and DNN towards the mean-field theory.
}Maximum mean-discrepancy $\mathrm{MMD^{2}}$ for a radial basis function
kernel with length scale $l=1/2$ \cite{Gretton12_723} between the
empirical distribution of scalar outputs $y_{\alpha}$ and the Gaussian
distribution with covariance matrix $K_{\alpha\beta}=\overline{C}{}_{\alpha,\beta}^{(A+1,A+1)}$
predicted by MFT \prettyref{eq:MFT_replicated}. Empirical $\mathrm{MMD^{2}}$
estimation across $2000$ realizations $(W,\xi)$. Average over $40$
realizations of $\{\bm{x}_{\alpha}\}_{\alpha=1,\ldots,10}$, $x_{\alpha,i}\stackrel{\text{i.i.d.}}{\sim}\protect\N(0,1)$
and $\text{dim}(\text{\ensuremath{\bm{x}}}_{\alpha})=4$ (error bars
showing standard error of the mean). ReLU activation $\phi(x)=\max(0,x)$.\textbf{
a) }$\mathrm{MMD^{2}}$ as a function of the width of the network
layer $n$ for $g^{2}=1.6$ and $g^{2}=1.2$ (inset), with $A=15$
and $\sigma^{2}=0.2$ and uncorrelated biases across time/layers.
\textbf{b) }$\mathrm{MMD^{2}}$ as a function of the depth or duration
$A$, for width $n=500$, $g^{2}=1.6$, and $\sigma^{2}=0.2$ and
uncorrelated biases. \textbf{c)} Same as a) but for biases that are
static across time/layers. \textbf{d)} Same as b) but for the static
bias case.\label{fig:Convergence-of-RNN-DNN}}
\end{figure}

Coming back to the general case, we next ask if the different off-diagonal
elements of the mean-field equations for RNN and DNN have observable
consequences. The answer is no if a linear readout is taken at a single
time point or layer $A$, correspondingly (cf.~\prettyref{eq:output_layer}
for the readout): This is a direct consequence of the identical diagonal
elements of the covariance $\overline{C}_{\alpha,\beta}^{(a,a)}$,
so that the predictive distribution \prettyref{eq:predictive_dist_final}
for the RNN and the DNN is identical in mean-field theory; the two
architectures have the same Gram matrix $K_{\alpha\beta}=\overline{C}{}_{\alpha,\beta}^{(A+1,A+1)}$
and thus the same predictive distribution \prettyref{eq:predictive_dist_final}.
This means that the two architectures have identical computational
capabilities in the limit of infinite layer width.

To check how quickly the mean-field theory is approached by finite-size
networks, we measure the maximum mean-discrepancy (MMD) \cite{Gretton12_723,Matthews18}
between the Gaussian distribution with covariance matrix $K_{\alpha\beta}$
and the empirical joint distribution of a set of scalar outputs $y_{\alpha}$,
\prettyref{eq:output_layer}, across realizations of $W$ and $\xi$.
The inputs $\bx_{\alpha}$ are random patterns presented to the first
layer or time step, respectively. We find that convergence is rather
fast for both architectures (\prettyref{fig:Convergence-of-RNN-DNN}a
and c). For sufficiently deep architectures $A\gg1$ as well as both
uncorrelated and static biases, RNNs systematically show a slower
convergence than DNNs, which could be anticipated due to the smaller
number of independently drawn Gaussian weights, $N^{2}$ versus $A\,N^{2}$.
This observation is in line with the MMD being larger for the RNN
than for the DNN for $A\gtrsim15$ (\prettyref{fig:Convergence-of-RNN-DNN}b
and d). This is also consistent with the coherent interplay of shared
connectivity and correlated activity across time steps in the RNN
(see \prettyref{app:RNN_instability}, \prettyref{fig:finite_size_effect}).
Overall, we find a faster convergence for uncorrelated biases than
for biases that are static over time or layers, respectively.

The temporal correlations present in RNNs become relevant in the case
of sequence processing. In such a setting, the network in each time
step $a$ receives a time dependent input $\bm{x}_{\alpha}^{(a)}$
with a non-trivial temporal correlation structure ${\bf x}_{\alpha}^{(a+\tau)\T}{\bf x}_{\beta}^{(a)}$
that drives the temporal correlations $\overline{C}{}_{\alpha,\beta}^{(a^{\prime}+\tau,a^{\prime})}$
of the RNN activations for $a^{\prime}\ge a$, see \prettyref{eq:MFT_replicated}.
If the latter are read out in each time step, temporal correlations
enter the kernel and thus influence task performance.

We finally note that we here use a separate readout layer. The realization
of readout weights as independent Gaussian variables causes vanishing
temporal correlations between the readouts and the activity in previous
layers or time steps, respectively. For the Gaussian kernel, however,
the presence or absence of a readout layer does not make any difference.
Alternatively, the readout of $n_{\mathrm{out}}$ signals could be
taken from an arbitrary choice of $n_{\mathrm{out}}$ neurons in the
last layer or time step, respectively, leading to the same kernel.

\subsection{Next-to-leading order corrections}

The saddle point approximation finds the dominant value for the correlation
$C^{(a,b)}$ by computing the stationary point of the action $\mathcal{S}_{\mathrm{aux}}$
given by \prettyref{eq:saddle_action}. A standard way to go beyond
the leading order is to obtain corrections of order $\mathcal{O}(n^{-1})$
by computing the Gaussian fluctuations of $C$ and $\tilde{C}$ around
their saddle-point values. To this end, we need the Hessian matrix
$\mathcal{S}_{\mathrm{aux},ij}^{(a,b),(c,d)}$, i.e., the derivatives
of $\mathcal{S}_{\mathrm{aux}}$ with respect to all combinations
of $C^{(a,b)}$ and $\tilde{C}^{(c,d)}$. The computation is global
in the sense that we consider the Hessian for all time points or layers,
$(a,b)$, $(c,d)$, simultaneously. This is in contrast to existing
perturbative approaches in DNNs, which proceed layer by layer \cite{ZavatoneVeth21_NeurIPS_II,Naveh21_NeurIPS}.
The negative inverse of the Hessian is the covariance matrix (or \emph{propagator})
$\Delta$,
\begin{eqnarray}
-\sum_{k=1}^{2}\,\mathcal{S}_{\mathrm{aux},ik}\,\Delta_{kj} & = & \delta_{i,j}\,\mathbf{1},\label{eq:S_2_Delta_inv}
\end{eqnarray}
which is meant as a 2x2 tensor equation consisting of tensors with
four time indices (see \prettyref{app:Next-to-leading-order-correction}
for details).

Due to the normalization $\mathcal{W}_{\text{aux}}(0\,|\,C)=0$, given
by \eqref{eq:W_rec_RNN} or \eqref{eq:W_rec_DNN}, respectively, the
Hessian has a zero in the upper left corner $\mathcal{S}_{\mathrm{aux},11}^{(a,b),(c,d)}\equiv\partial^{2}\mathcal{S}_{\mathrm{aux}}/\partial C^{(a,b)}\partial C^{(c,d)}\equiv0$.
This implies a zero in the lower right corner of its inverse $\Delta$.
As a consequence, the off-diagonal elements $\Delta_{12}^{(a,b),(c,d)}=\Delta_{21}^{(c,d),(a,b)}$,
which are response functions and time-reversed (transposed) of one
another, can be determined independently of $\Delta_{11}$ through
\begin{eqnarray*}
-\mathcal{S}_{\mathrm{aux},12}\,\Delta_{21} & = & \mathbf{1}.
\end{eqnarray*}
A direct calculation (\prettyref{app:Next-to-leading-order-correction})
shows that for DNN and for RNN their equal-time entries $\Delta_{12}^{(a,a),(e,e)}$
are causal, i.e., they vanish for $a<e$, and are else given by
\begin{eqnarray}
\Delta_{12}^{(a,a),(e,e)}=\langle C^{(a,a)}\,\tilde{C}^{(e,e)}\rangle & = & n^{-1}\,\nu_{e-1}^{-1}\,\prod_{k=e}^{a-1}F_{k},\label{eq:response_function_main}
\end{eqnarray}
where $F_{k}=g_{k+1}^{2}\,\partial\,\langle\phi^{(k)}\phi^{(k)}\rangle/\partial C^{(k,k)}$
is the linear response of the correlator $g_{k+1}^{2}\langle\phi^{(k)}\phi^{(k)}\rangle$
in layer $k$ with regard to changes of the variance $C^{(k,k)}$
of the latent variables in that layer; for RNN we have $\nu=1$. With
the ReLU non-linearity, $F_{k}$ takes the particularly simple form
$F_{k}=\frac{1}{2}g_{k+1}^{2}$. This implies that the response functions
show exponentially decaying behavior for $g_{k}^{2}<2$, which is
the stable regime of the mean-field equations.

The response function $\Delta_{12}$ describes how the variability
$\delta C$ of the variance in the first layer or time step $0$ is
propagated to a downstream layer or later time step $a$. Since the
joint statistics of $C$ and $\tilde{C}$ follows $C,\tilde{C}\sim\exp\left(\mathcal{S}_{\mathrm{aux}}(C,\tilde{C})\right)$,
to linear order the variability $\delta C$ in layer $0$ affects
the variability in layer $a$ as

\begin{equation}
C^{(a,a)}=\langle C^{(a,a)}\rangle_{C^{(0,0)}=\bar{C}^{(0,0)}}+\Delta_{12}^{(a,a),(0,0)}\,\delta C^{(0,0)}+\mathcal{O}(\delta^{2}),\label{eq:input_deviation}
\end{equation}
where we used that $\partial\langle C^{(a,a)}\rangle_{C^{(0,0)}}/\partial C^{(0,0)}=\langle C^{(a,a)}\tilde{C}^{(0,0)}\rangle=\Delta_{12}^{(a,a),(0,0)}$.
This can also be seen by noting that the solution \eqref{eq:response_function_main}
for $\Delta_{12}$ is identical to the linear response of the iterative
mean-field equations \eqref{eq:MFT} with regard to an infinitesimal
perturbation of $C^{(e,e)}$.

The diagonal element $\Delta_{11}^{(a,a),(a,a)}$ of the propagator
in \eqref{eq:S_2_Delta_inv} is the variability of the variance in
layer $a$. Considering the other entries of \eqref{eq:S_2_Delta_inv},
we find that it obeys (see also \eqref{eq:cov_from_prop})
\begin{eqnarray}
\Delta_{11} & = & \Delta_{12}\,\mathcal{S}_{\mathrm{aux},22}\,\Delta_{21},\label{eq:Delta_11}
\end{eqnarray}
where in\begin{eqnarray*}\mathcal{S}_{\mathrm{aux},22}^{(a,b),(c,d)}&=&\frac{\partial^{2}\mathcal{S}_{\mathrm{aux}}}{\partial\tilde{C}^{(a,b)}\,\partial\tilde{C}^{(c,d)}}\\&=&n\,\Big\langle\phi^{(a-1)}\phi^{(b-1)},\phi^{(c-1)}\phi^{(d-1)}\Big\rangle_{c}\\
&& \times \cases{g_{a}^{4}\,\nu_{a-1}\,\delta_{a,b}\,\delta_{c,d}\,\delta_{a,c} & \text{DNN}\\g^{4} & \text{RNN}}\\&=:&n\,G_{a-1,b-1,c-1,d-1},\end{eqnarray*}the connected correlation function (second cumulant) $\langle\circ,\circ\rangle_{c}$
of $\phi\phi$ appears in the second line. The expression \eqref{eq:Delta_11}
shows that fluctuations generated by $\mathcal{S}_{\mathrm{aux},22}^{(2)}$
propagate through the network forward in time or layer, expressed
by the two factors $\Delta_{12}$ and $\Delta_{21}$. The Kronecker
$\delta$ in the expression for the DNN imply that fluctuations are
intrinsically-generated only within layers, whereas in the RNN fluctuations
between different time points are correlated due to the weight sharing.
One is typically interested in the fluctuations measured in a single
layer $a$, for example in the readout layer. In this case due to
$\Delta_{12}^{(a,a),(c,d)}\propto\delta_{c,d}$ and by \eqref{eq:Delta_11},
one only needs $\mathcal{S}_{\mathrm{aux},22}^{(a,a),(c,c)}$ to obtain\begin{eqnarray}\Delta_{11}^{(a,a),(a,a)}&=&n^{-1}\,\cases{\,\sum_{a'=1}^{a}\left\{ \prod_{k=a'}^{a-1}F_{k}^{2}\right\} \nu_{a'-1}^{-1}\,G_{a^{\prime}-1,a^{\prime}-1,a^{\prime}-1,a^{\prime}-1}\\\,\sum_{a^{\prime},c^{\prime}=1}^{a}\Bigg\{\prod_{k=a'}^{a-1}F_{k}\Bigg\}\,\Bigg\{\prod_{l=c'}^{a-1}F_{l}\Bigg\}\,G_{a^{\prime}-1,a^{\prime}-1,c^{\prime}-1,c^{\prime}-1}} \label{eq:Delta_11_maintext} \end{eqnarray}where
the upper result holds for the DNN and the lower for the RNN. The
two expressions differ by the presence of the additional summation
index $c^{\prime}$ in the case of the RNN and by the appearance of
the relative layer sizes $\nu$ for the DNN. The reason is, again,
that correlated fluctuations in the RNN are generated also across
different time steps due to weight sharing. A further difference is
that $F$ and $G$ must be evaluated at the corresponding mean-field
solutions of the two network architectures. For the ReLU activation
function we find $G_{a-1,a-1,a-1,a-1}=\frac{5}{4}g_{a}^{4}\,C^{(a-1,a-1)}C^{(a-1,a-1)}$
for the DNN as well as $G_{a-1,a-1,c-1,c-1}=g^{4}\,\Big(\langle\phi^{(a-1)}\phi^{(a-1)}\phi^{(c-1)}\phi^{(c-1)}\rangle-\frac{1}{4}C^{(a-1,a-1)}C^{(c-1,c-1)}\Big)$
with $\langle\phi^{(a-1)}\phi^{(a-1)}\phi^{(c-1)}\phi^{(c-1)}\rangle$
known from \cite{Cho09} for the RNN (see \prettyref{eq:quartic_relu_average}).
\begin{figure}
\begin{centering}
\includegraphics{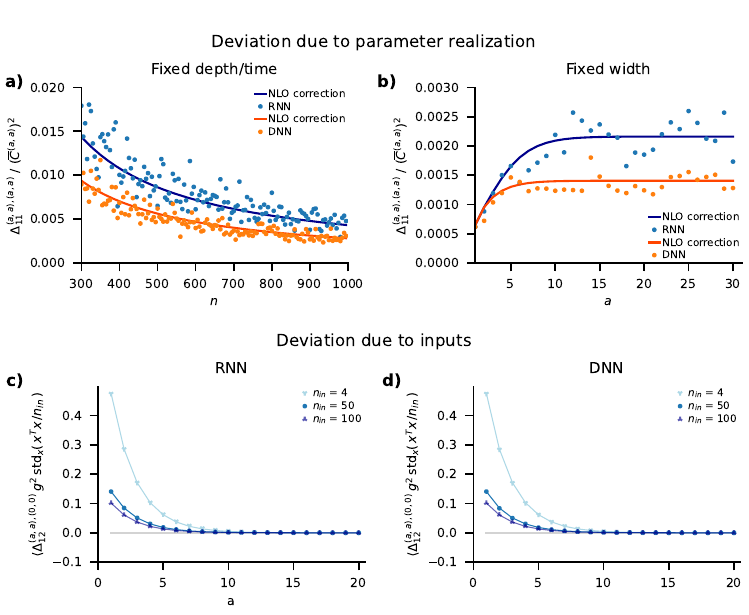}
\par\end{centering}
\caption{\textbf{NLO corrections for DNN and RNN. }Fluctuations $\Delta_{11}^{(a,a),(a,a)}$
of the covariance $C^{(a,a)}$ around its mean value $\overline{C}^{(a,a)}$
in the last layer $a=A+1$ across realizations of weights $W$ and
biases $\xi$.\textbf{ a)} $\Delta^{(a,a),(a,a)}$ for $a=A+1$ as
a function of the network width $n$; simulations as dots, estimated
across $100$ realizations of $W$ and $\xi$, and theoretical prediction
\eqref{eq:Delta_11_maintext} as curves for RNN (blue) and DNN (orange).
Network parameters: $g^{2}=g_{a}^{2}=1.2$, $\sigma^{2}=\sigma_{a}^{2}=0.2$,
$A=15$. All realizations use the same fixed input $\boldsymbol{x}\in\mathbb{R}^{4}$
with $x_{i}\stackrel{\text{i.i.d.}}{\sim}\protect\N(0,1)$ and ReLU
activation $\phi(x)=\max(0,x)$. \textbf{b) }$\Delta_{11}^{(a,a),(a,a)}$
as a function of the network depth or readout time $a$, respectively;
simulations as dots, estimated from $100$ realizations of $W$ and
$\xi$, and theoretical prediction \eqref{eq:Delta_11_maintext}
as curves for RNN (blue) and DNN (orange). Network parameters: $g^{2}=g_{a}^{2}=1.2$,
$\sigma^{2}=\sigma_{a}^{2}=0.2$ and $n=2000$. Same input $\boldsymbol{x}\in\mathbb{R}^{4}$
with $x_{i}\stackrel{\text{i.i.d.}}{\sim}\protect\N(0,1)$ and ReLU
activation $\phi(x)=\max(0,x)$ as in panel a).\label{fig:Fluctuations-corrections}
\textbf{c) }RNN, \textbf{d)} DNN fluctuations of the mean field covariance
induced by input fluctuations as a function of the layer / time step
$a$. Dots represent the simulation result obtained as the standard
deviation with respect to different input realizations $x$ of the
weight averaged overlaps $\text{\ensuremath{\overline{C}^{(a,a)}}(x)=\ensuremath{\big\langle\frac{1}{n_{a}}\phi^{(a-1)\protect\T}\phi{}^{(a-1)}\big\rangle_{W}}(x)}$;
we used $100$ network realizations in the simulations. Lines represent
the theoretical prediction based on linear response theory \eqref{eq:input_deviation}.
Network parameters: $g^{2}=g_{a}^{2}=1.2$, $\sigma^{2}=\sigma_{a}^{2}=0.2$
and $n=500$.}
\end{figure}

The comparison of the theoretical prediction \eqref{eq:Delta_11_maintext}
to an estimate of the fluctuations of $C$ is shown in \prettyref{fig:Fluctuations-corrections}.
As expected for the next-to-leading order corrections being $\propto\mathcal{O}(n^{-1})$,
the variability declines inversely proportional to $n$, as shown
in \prettyref{fig:Fluctuations-corrections}a. The variability for
the RNN is throughout larger than for the DNN. This can be anticipated
from the expression \eqref{eq:Delta_11_maintext}, which shows that
for the RNN fluctuations of $\phi\phi$ across different time-steps
$a^{\prime}\neq c^{\prime}$ drive fluctuations at the later final
time point, while for the DNN only fluctuations from within the same
layer $a^{\prime}$ are propagated forward to the final layer. As
a function of depth (DNN) or time step (RNN), respectively, the fluctuation
corrections show a characteristic form with an initial increase followed
by a plateau, shown in \prettyref{fig:Fluctuations-corrections}b.
This is due to the the exponential decay of the causal response functions
$\Delta_{12}^{(a,a),(a^{\prime},a^{\prime})}$ with the distance $a-a^{\prime}$
in \eqref{eq:Delta_11_maintext}. For a stationary mean-field solution
one would have $F_{k}=F$, so the depth or time scale is $\tau=-\big(\ln F\big)^{-1}$,
because $\Delta_{12}^{(a,a),(a^{\prime},a^{\prime})}\propto F^{a-a^{\prime}}=e^{(a-a^{\prime})\,\ln F}=e^{-\frac{a-a^{\prime}}{\tau}}$.
For ReLU activations this evaluates to $\tau=\big(\ln2-2\ln g\big)^{-1}\simeq2$
for $g^{2}=1.2$. As expected, at the point $g^{2}=2$, where the
mean-field solution looses stability, also this time scale diverges.
The function $\Delta_{12}^{2}$ appearing in \eqref{eq:Delta_11_maintext}
thus declines with about a unit scale for the given parameters. The
initial increase of fluctuations $\Delta_{11}$ results from the convolution
with $\Delta_{12}^{2}$ of the variance of $\phi\phi$, which itself
has a transient behavior inherited from the transient of the mean-field
solution shown in \prettyref{fig:Mean-field-theory-for-DNN-RNN}.
At large depth, fluctuations saturate on a plateau, the height of
which is given by the accumulated fluctuations of the previous layers,
discounted by $\Delta_{12}^{2}$; the driving fluctuations of $\phi\phi$
in this limit become constant as the mean-field solution $\bar{C}$
approaches its stationary plateau. Since $\Delta_{12}$ has the same
form for the RNN and the DNN, analog considerations explain the rise-and-plateau
shape for the RNN.

\prettyref{fig:Fluctuations-corrections}c and d show the propagation
of fluctuations of the input through the layers of the DNN (c) and
through time in the RNN (d). The simulations (dots) show good agreement
with the theoretical prediction from linear response theory \eqref{eq:input_deviation}
(curves) for both architectures. In fact, the theoretical predictions
for the statistics within a layer (DNN) and for equal times (RNN)
are identical. One observes the exponential decay of the input variability
with deeper layers or later time, respectively, as pointed out above.

The fluctuations $\Delta_{11}$ yield a non-Gaussian distribution
of the output of the network
\begin{eqnarray*}
p(\bm{y}\,|\,\bm{x}) & \simeq & \int d\tilde{\bm{y}}\,\exp\left(\tilde{\bm{y}}^{\T}\bm{y}\right)\,\Big\langle\exp\left(\frac{1}{2}\tilde{\bm{y}}^{\T}C\tilde{\bm{y}}\right)\Big\rangle_{C\sim\N(\bar{C},\Delta_{11})}\\
 & = & \int d\tilde{\bm{y}}\,\exp\left(\tilde{\bm{y}}^{\T}\bm{y}\right)\,\exp\Big(\frac{1}{2}\tilde{\bm{y}}^{\T}\overline{C}{}^{(A+1,A+1)}\tilde{\bm{y}}\\
 &  & \phantom{\int d\tilde{\bm{y}}\,\exp\left(\tilde{\bm{y}}^{\T}\bm{y}\right)\,\exp\Big(}+\frac{1}{8}\,\Delta_{11}^{(A+1,A+1),(A+1,A+1)}\,(\tilde{\bm{y}}^{\T}\tilde{\bm{y}})^{2}\Big).
\end{eqnarray*}
The latter term $\frac{1}{8}\,\Delta_{11}^{(A+1,A+1),(A+1,A+1)}\,(\tilde{\bm{y}}^{\T}\tilde{\bm{y}})^{2}$
describes forth order cumulants of the output $y$ between pairs of
indices $i,j$ due to the appearance of $(\tilde{\bm{y}}^{\T}\tilde{\bm{y}})^{2}=\sum_{i,j}\tilde{y}_{i}^{2}\tilde{y}_{j}^{2}$.
This approximation may be combined with results by Zavatone-Veth et
al. \cite{ZavatoneVeth21_NeurIPS_II} or Naveh and Ringel \cite{Naveh21_NeurIPS}
to obtain finite-size corrections for the statistics of the predictive
distribution, extending their results to networks of arbitrary depth
and to arbitrary activation functions.

\section{Discussion}

We present a unified derivation of the mean-field theory for deep
(DNN) and recurrent neural networks (RNN) with arbitrary activation
functions using field-theoretical methods. The derivation in particular
yields the Gaussian process kernel that predicts the performance of
networks trained in a Bayesian way. For the network priors we furthermore
present explicit next-to-leading-order corrections to the mean-field
theories, which are valid for general activation functions.

The mean-field theories for the statistics within a layer of the DNN
and for the equal-time statistics of the RNN are identical, even if
temporally correlated input sequences are supplied to the latter network.
The reason is that the mean-field equations (\ref{eq:MFT_replicated})
form a closed system of equations for this subset of the statistics;
they can be solved independently of the correlations across time or
layers, respectively. This justifies the `annealed approximation'
\cite{Fischer91} for RNNs where the couplings are redrawn at each
time step—which corresponds to the DNN-prior. It is also compatible
with earlier work \cite{Chen18_873} which compares simulations of
networks with tied weights (RNN) to the mean-field theory for untied
weights (DNN). Intriguingly, the equivalence of the equal-time statistics
implies that the predictive distributions $p(\bm{y}^{*}\,|\,\bm{x}^{*},\bm{Y},\bm{X})$
of DNNs and RNNs are identical \cite{Hron20_10541}, given the readout
is taken only from the final layer or the last time step, respectively.

There are qualitative differences between the mean-field theories
for the correlations across time in the RNN and across layers of the
DNN: Correlations across layers vanish in the DNN, while the weight
sharing in the RNN generally causes non-trivial correlations across
time. For point-symmetric activation functions, these correlations
also vanish in the RNN if the bias is absent (or uncorrelated across
time steps) and the input is provided only in the first step.  In
general, a linear readout from activations that are taken across different
time points or layers, respectively, yields different Gaussian process
kernels for the RNN compared to the DNN. Even if the readout is taken
at a single layer or time point, respectively, there is an observable
difference between DNN and RNN when the fluctuation corrections to
the mean field kernels are taken into account: While in RNNs fluctuations
are in general correlated across time steps, due to the weight sharing
between the time points, fluctuations are independent across layers
for the DNN. As a result, in case of the RNN, the fluctuations at
the readout $\Delta_{11}^{(A+1,A+1),(A+1,A+1)}$ depend on the fluctuations
between all combinations of time points, whereas for the DNN only
within-layer fluctuations from all previous layers influence the result.

Numerically, the convergence of finite-size networks of both architectures
to the mean-field theory is generally fast. The RNN converges typically
slower than the DNN, at least for long times and correspondingly deep
networks. We hypothesize that the temporally correlated activity in
the RNN is the cause: The realization of the coupling matrix is the
same for all time steps. Also, fluctuations of the activity are coherent
over time. Activity and connectivity therefore interact coherently
over multiple time steps, so that variations of the connectivity across
realizations may cause a corresponding variability of activity. In
a DNN, in contrast, both activity and connectivity are uncorrelated
across layers, so that variations due to different realizations of
the couplings average out over layers. The larger discrepancy between
the theoretical prediction and the simulation in case of the RNN as
compared to the DNN can also be observed for the next-to-leading-order
corrections. Although the form of the corrections differ between RNN
and DNN, for the RNN the range of widths that the theory can capture
within some given error margin is smaller than the range of validity
for the DNN. This implies that overall finite size effects are more
relevant in RNNs than they are in DNNs.

Identical mean-field theories in the single-input case and for point-symmetric
activation functions were already presented in ref.~\cite{Mozeika20_168301}
in the context of a characterization of the space of Boolean functions
implemented by randomly coupled DNNs and RNNs. Since our work differs
on a conceptual level, the implications of the results differ: In
the Bayesian inference picture, the equivalent mean-field theories
imply identical performance of the networks for both architectures
at large width; for the characterization of computed Boolean functions,
the equivalent mean-field theories imply an equivalent set of functions
implemented by any two random instances of the two architectures at
large width. The conceptual difference leads to further differences
on the technical level: The inputs and outputs considered here include
analog values and they are presented not only to the first layer or
time step, respectively, but also in a sequential manner at subsequent
times or layers. Finally, the disorder average plays a subtle but
fundamentally different role in the two works: In ref.~\cite{Mozeika20_168301},
the disorder average extracts the typical behavior of any single,
sufficiently large, instance of a randomly coupled network. In contrast,
in the Bayesian framework considered in this manuscript, the disorder
average naturally arises from the marginalization of the parameter
prior, i.e., one here considers ensembles of random networks.

The analysis of RNNs and DNNs in this manuscript is based on methods
from statistical field theory and our results are formulated in that
language \cite{ZinnJustin96}. It is, however, worth noting that
we are actually dealing with multi-variate random variables in $n$
dimensions rather than fields, so that mathematical complications
with regard to the latter do not appear here. Moreover, the field-theoretical
approach to compute the leading order as a saddle point approximation
and the next-to-leading order in $n^{-1}$ from the stability matrix
can be connected to the approach of large-deviation theory \cite{vanMeegen21_158302};
exploring this link further is an interesting topic for future research.

There has been previous theoretical work on networks of finite width
$n_{\ell}<\infty$ that is, however, restricted to DNNs. Two different
approximation techniques have been employed. The perturbative approach
computes corrections where the non-linear terms constitute the expansion
parameter; this typically requires analytic activation functions that
can be expanded in low order monomials. The Edgeworth expansion, in
contrast, obtains approximations in the strength of the non-Gaussian
cumulants as an expansion parameter.~Refs.~\cite{Yaida20,Dyer20_ICLR,Halverson21_035002,Aitken20_06687,Roberts22}
have presented approaches based on perturbation theory, while refs.~\cite{Antognini19_arxiv,Naveh21_064301}
employed an Edgeworth expansion. These corrections were computed
either in the framework of Bayesian inference \cite{Yaida20,Antognini19_arxiv,Naveh21_064301,Roberts22}
or gradient-based training \cite{Dyer20_ICLR,Huang20_4542,Aitken20_06687,Roberts22}.
The dynamics of the neural-tangent kernel for deep networks with finite
width has been studied in ref.~\cite{Huang20_4542}. For specific
deep networks of finite width with linear or ReLU activation functions
the single-input prior was computed exactly in terms of the Meijer
G function in refs.~\cite{ZavatoneVeth21_NeurIPS_I,Noci21_DTA7Bgrai-Q}.

In this manuscript, we considered the output dimension $n_{\mathrm{out}}$
and the depth or number of time-steps $A$ to be fixed. Other works
\cite{Roberts22,Grosvenor22_81} investigated the limit $n\to\infty$
where $A/n$ is small but finite. Due to the exponential decay of
the response functions $\Delta_{12}$, in our setup, this distinction
becomes irrelevant as soon as one considers enough layers or time-steps
such that the statistics are stationary. In contrast, the output dimension
needs to be $O(1)$ in our setup; otherwise, the exponent $\frac{1}{2}\tilde{\bm{y}}^{\T}C^{(A+1,A+1)}\tilde{\bm{y}}$
in \prettyref{eq:single_output_prob} would not be $O(1)$ and the
term would need to be taken into account when deriving the mean-field
equation \eqref{eq:MFT} via \prettyref{eq:saddle_action}. Even if
the output dimension is $O(1)$, this generally implies a $O(\frac{1}{n})$
correction to the mean-field equations.

While we here focus on the network prior, the recent works by Zavatone-Veth
et al.~\cite{ZavatoneVeth21_NeurIPS_II} and Naveh and Ringel \cite{Naveh21_NeurIPS}
directly address perturbative effects on the predictive distribution.
Zavatone-Veth et al.~conjecture a general form of finite width corrections
for network observables which result from the linear readout layer
and the quadratic loss function. Their expression requires the knowledge
of $\mathrm{cov}_{\mathcal{W}}(O,K)$, the covariance of the kernel
$K$ and the considered observable $O$ across realizations of the
feature map parameters $\mathcal{W}$. Explicit finite-width corrections
are obtained for fully connected and convolutional deep linear networks
with and without skip connections. Their computation is perturbative
in the non-Gaussian terms and proceeds iteratively, layer by layer.
The work by Naveh and Ringel consists of two parts: First, they compute
the correction to the mean of the predictive distribution that arises
from non-Gaussian cumulants of the process. The correction follows
as the solution of a set of self-consistency equations. Second, they
compute non-Gaussian corrections perturbatively for shallow fully
connected and convolutional feed-forward networks with linear and
quadratic activation functions. The general results by Zavatone-Veth
et al.~and Naveh and Ringel could directly be applied to compute
corrections to the mean of the predictive distribution based on the
finite-size corrections we have found here, thereby extending their
work to nonlinear DNNs and RNNs with arbitrary activation functions.

In a concurrent work, Grosvenor and Jefferson \cite{Grosvenor22_81}
extend the dynamical mean-field theory of recurrent networks in continuous
time \cite{Sompolinsky88_259} within the field-theoretical formulation
\cite{Crisanti18_062120,Helias20_970} beyond the mean-field limit
in order to obtain finite-size corrections for stationary statistics
and for the transition to chaos. They proceed with a perturbation
expansion around the infinite width mean-field solution, taking into
account the nonlinearities approximately via their low-order Taylor
monomials and find that the width over the depth is the effective
expansion parameter. In contrast to our work, in addition to the different
dynamical network equations, they focus on the stationary regime whereas
we take the non-stationary propagation of the input through layers
or time, respectively, into account.

A further difference between these previous works and the present
work is that we present the next-to-leading order correction ($\propto n^{-1})$
in the fluctuation expansion of the auxiliary fields, as opposed to
a perturbation expansion on the level of neuronal fields. We obtain
these results in a global manner for all layers or time steps simultaneously,
rather than iteratively across layers. Thus, we do not need to truncate
the approximation at intermediate layers. Moreover, our approach applies
to deep and recurrent fully connected networks, including the full
nonlinearity, rather than its polynomial approximation. To our knowledge,
our work is the first to derive mean-field and beyond mean-field corrections
for DNN and RNN in a unified framework which allows us to discuss
the qualitative differences between these two architectures.

\ack{}{We would like to thank Bo Li, Alexander Mozeika, and David Saad for
bringing their related work to our attention. Furthermore, we would
like to thank the anonymous reviewers for helpful comments and suggestions.
This work was partially supported by the European Union’s Horizon
2020 research and innovation program under Grant agreement No. 945539
(Human Brain Project SGA3), the Helmholtz Association Initiative and
Networking Fund under project number SO-092 (Advanced Computing Architectures,
ACA), the German Federal Ministry for Education and Research (BMBF
Grant 01IS19077A), and the Excellence Initiative of the German federal
and state governments (ERS PF-JARA-SDS005).}

\section{Appendix\label{app:Appendix}}

\subsection{Unified field theoretical approach for multiple input sequences\label{app:replicated-Field-theory-DNN-RNN}}

Here, we show the derivation of the mean-field equations with more
than one input sequence $\{\bm{x}_{\alpha}^{(0)},\dots,\bm{x}_{\alpha}^{(A)}\}$,
the generalization of the derivation presented in the main text. We
introduce Greek indices $\alpha\in\{1,\dots,n_{D}\}$ for the different
input vectors that we also call `replicas' in the following. Equations
for the single-replicon case in the main text can be obtained by setting
$n_{D}=1$; the non-sequential input case follows by setting $\bm{x}_{\alpha}^{(a)}=0$
for $a>0$ and all $\alpha$.

\subsubsection{Action and auxiliary variables}

We start from the parameterized likelihood for multiple replicas

\begin{eqnarray*}
p(\bm{Y}\,|\,\bm{X},\bm{\theta}) & =\prod_{\alpha=1}^{n_{D}} & \Bigg\{\int\D\bm{h}_{\alpha}\,\delta\text{\ensuremath{\left(\bm{y}_{\alpha}-\bm{W}^{(\text{out})}\bm{\phi}_{\alpha}^{(A)}-\bm{\xi}^{(A+1)}\right)}}\\
 &  & \times\prod_{a=1}^{A}\delta\left(\bm{h}_{\alpha}^{(a)}-\bm{W}^{(a)}\bm{\phi}_{\alpha}^{(a-1)}-\bm{W}^{(\text{in},a)}\bm{x}_{\alpha}^{(a)}-\bm{\xi}^{(a)}\right)\\
 &  & \times\delta\left(\bm{h}_{\alpha}^{(0)}-\bm{W}^{(\text{in},0)}\bm{x}_{\alpha}^{(0)}-\bm{\xi}^{(0)}\right)\Bigg\}.
\end{eqnarray*}
Expressing the Dirac distributions as integrals $\delta(x)=\int_{i\mathbb{R}}\frac{d\tilde{x}}{2\pi i}\,e^{\tilde{x}\,x}$,
we obtain for the network prior $p(\bm{Y}\,|\,\bm{X})=\int d\bm{\theta}\,p(\bm{Y}\,|\,\bm{X},\bm{\theta})\,p(\bm{\theta})$
the expression

\begin{eqnarray}
p(\bm{Y}\,|\,\bm{X}) & = & \prod_{\alpha=1}^{n_{D}}\left\{ \int d\tilde{\bm{y}}_{\alpha}\int D\bm{h}_{\alpha}\int D\tilde{\bm{h}}_{\alpha}\right\} \,e^{\tilde{y}_{i,\alpha}y_{i,\alpha}+\sum_{a=0}^{A}\tilde{h}_{i,\alpha}^{(a)}\,h_{i,\alpha}^{(a)}}\nonumber \\
 & \quad & \times\langle e^{-\tilde{y}_{i,\alpha}W_{ij}^{(\text{out)}}\phi_{j,\alpha}^{(A)}}\rangle_{\bm{W}^{(\text{out})}}\:\langle e^{-\sum_{a=1}^{A}\tilde{h}_{i,\alpha}^{(a)}W_{ij}^{(a)}\phi_{j,\alpha}^{(a-1)}}\rangle_{\{\bm{W}^{(a)}\}}\nonumber \\
 & \quad & \times\langle e^{-\sum_{a=0}^{A}\tilde{h}_{i,\alpha}^{(a)}W_{ij}^{(\text{in},a)}x_{j,\alpha}^{(a)}}\rangle_{\{\bm{W}^{(\text{in},a)}\}}\nonumber \\
 & \quad & \times\langle e^{-\sum_{\alpha=1}^{n_{D}}\sum_{a=0}^{A}\tilde{h}_{i,\alpha}^{(a)}\xi_{i}^{(a)}-\sum_{\alpha=1}^{n_{D}}\tilde{y}_{i,\alpha}\xi_{i}^{(A+1)}}\rangle_{\{\bm{\xi}^{(a)}\}}.\label{eq:gen_mom_gen_func}
\end{eqnarray}
Here, and throughout this section, we use an implicit summation convention
for lower indices that appear twice in the exponent, e.g., $\tilde{y}_{i,\alpha}y_{i,\alpha}\equiv\sum_{\alpha=1}^{n_{D}}\sum_{i=1}^{n_{\mathrm{out}}}\tilde{y}_{i,\alpha}y_{i,\alpha}$,
but write the sum over time steps explicitly to avoid ambiguities
in their limits. Note that for the $\DNN$, the number of neurons
per layer can differ such that formally the upper limits of the implicit
sums over neuron indices $i$ or $j$ depends on the layer index $a$.
We also used the independence of the different weight matrices and
biases to obtain factorizing expectation values in \prettyref{eq:gen_mom_gen_func}.

In the following, we compute these expectation values separately,
starting with the output weights and biases. These are independent
across neurons and we obtain
\[
\Big\langle\exp\left(-\sum_{\alpha=1}^{n_{D}}\tilde{y}_{i,\alpha}\xi_{i}^{(A+1)}\right)\Big\rangle_{\bm{\xi}^{(A+1)}}=\exp\left(\frac{\sigma_{A+1}^{2}}{2}\sum_{\alpha,\beta=1}^{n_{D}}\tilde{y}_{i,\alpha}\tilde{y}_{i,\beta}\right),
\]

\[
\Big\langle\exp\left(-\tilde{y}_{i,\alpha}W_{ij}^{(\text{out)}}\phi_{j,\alpha}^{(A)}\right)\Big\rangle_{\bm{W}^{(\text{out})}}=\exp\left(\frac{g_{A+1}^{2}}{2n_{A}}\tilde{y}_{i,\alpha}\phi_{j,\alpha}^{(A)}\phi_{j,\beta}^{(A)}\tilde{y}_{i,\beta}\right).
\]
Now, we calculate the respective averages for the RNN and DNN separately.
For a RNN, the weight sharing $\bm{W}^{(a)}\equiv\bm{W}$ across time
steps $a$ leads to a double sum $\sum_{a,b}$ appearing in the average
over the recurrent part
\begin{eqnarray*}
\Big\langle\exp & \left(-\sum_{a=1}^{A}\tilde{h}_{i,\alpha}^{(a)}W_{ij}\phi_{j,\alpha}^{(a-1)}\right)\Big\rangle_{\bm{W}}\\
 & =\exp\left(\frac{1}{2}\sum_{a,b=1}^{A}\frac{g^{2}}{n}{\textstyle \tilde{h}_{i,\alpha}^{(a)}\phi_{j,\alpha}^{(a-1)}\phi_{j,\beta}^{(b-1)}\tilde{h}_{i,\beta}^{(b)}}\right) & ,\quad\mathrm{RNN}.
\end{eqnarray*}
 In contrast, for a DNN, the analogous calculation leads to a single
sum $\sum_{a}$ 
\begin{eqnarray*}
\Big\langle\exp & \left(-\sum_{a=1}^{A}\tilde{h}_{i,\alpha}^{(a)}W_{ij}^{(a)}\phi_{j,\alpha}^{(a-1)}\right)\Big\rangle_{\{\bm{W}^{(a)}\}}\\
 & =\prod_{a=1}^{A}\Big\langle\exp\left(-\tilde{h}_{i,\alpha}^{(a)}W_{ij}^{(a)}\phi_{j,\alpha}^{(a-1)}\right)\Big\rangle_{\bm{W}^{(a)}}\\
 & =\exp\left(\frac{1}{2}\sum_{a=1}^{A}\frac{g_{a}^{2}}{n_{a-1}}\tilde{h}_{i,\alpha}^{(a)}\phi_{j,\alpha}^{(a-1)}\phi_{j,\beta}^{(a-1)}\tilde{h}_{i,\beta}^{(a)}\right) & ,\quad\mathrm{DNN}.
\end{eqnarray*}
The calculation for the inputs and biases is analogous; for the RNN
it yields
\begin{eqnarray*}
\Big\langle\exp & \left(-\sum_{a=0}^{A}\sum_{\alpha=1}^{n_{D}}\tilde{h}_{i,\alpha}^{(a)}\xi_{i}\right)\Big\rangle_{\bm{\xi}}\\
 & =\exp\left(\frac{\sigma^{2}}{2}\sum_{a,b=0}^{A}\sum_{\alpha,\beta=1}^{n_{D}}\tilde{h}_{i,\alpha}^{(a)}\tilde{h}_{i,\beta}^{(b)}\right), & \quad\mathrm{RNN},
\end{eqnarray*}

\begin{eqnarray*}
\Big\langle\exp & \left(-\sum_{a=0}^{A}\tilde{h}_{i,\alpha}^{(a)}W_{ij}^{(\text{in})}x_{j,\alpha}^{(a)}\right)\Big\rangle_{\bm{W}^{(\text{in})}}\\
 & =\exp\left(\frac{1}{2}\sum_{a,b=0}^{A}\frac{g_{\text{0}}^{2}}{n_{\text{in}}}{\textstyle \tilde{h}_{i,\alpha}^{(a)}x_{j,\alpha}^{(a)}x_{j,\beta}^{(b)}\tilde{h}_{i,\beta}^{(b)}}\right) & ,\quad\mathrm{RNN}.
\end{eqnarray*}
For the DNN, we get
\begin{eqnarray*}
\Big\langle\exp & \left(-\sum_{a=0}^{A}\sum_{\alpha=1}^{n_{D}}\tilde{h}_{i,\alpha}^{(a)}\xi_{i}^{(a)}\right)\Big\rangle_{\{\bm{\xi}^{(a)}\}}\\
 & =\prod_{a=0}^{A}\Big\langle\exp\left(-\sum_{\alpha=1}^{n_{D}}\tilde{h}_{i,\alpha}^{(a)}\xi_{i}^{(a)}\right)\Big\rangle_{\bm{\xi}^{(a)}}\\
 & =\exp\left(\frac{1}{2}\sum_{a=0}^{A}\sigma_{a}^{2}\sum_{\alpha,\beta=1}^{n_{D}}\tilde{h}_{i,\alpha}^{(a)}\tilde{h}_{i,\beta}^{(a)}\right), & \quad\mathrm{DNN},
\end{eqnarray*}
\begin{eqnarray*}
\Big\langle\exp & \left(-\sum_{a=0}^{A}\tilde{h}_{i,\alpha}^{(a)}W_{ij}^{(\text{in},a)}x_{j,\alpha}^{(a)}\right)\Big\rangle_{\{\bm{W}^{(\text{in},a)}\}}\\
 & =\prod_{a=0}^{A}\Big\langle\exp\left(-\tilde{h}_{i,\alpha}^{(a)}W_{ij}^{(\text{in},a)}x_{j,\alpha}^{(a)}\right)\Big\rangle_{\bm{W}^{(\text{in},a)}}\\
 & =\exp\left(\frac{1}{2}\sum_{a=0}^{A}\frac{g_{\text{0}}^{2}}{n_{\text{in}}}\tilde{h}_{i,\alpha}^{(a)}x_{j,\alpha}^{(a)}x_{j,\beta}^{(a)}\tilde{h}_{i,\beta}^{(a)}\right), & \quad\mathrm{DNN}.
\end{eqnarray*}
For the RNN, the replicas as well as the time steps are coupled by
the products $\phi_{j,\alpha}^{(a-1)}\phi_{j,\beta}^{(b-1)}$ and
$x_{j,\alpha}^{(a)}x_{j,\beta}^{(b)}$, while for the DNN only products
of terms within the same layer occur, $\phi_{j,\alpha}^{(a-1)}\phi_{j,\beta}^{(a-1)}$
and $x_{j,\alpha}^{(a)}x_{j,\beta}^{(a)}$. As we will show below,
this leads to different layers in the DNN being uncorrelated, while
different time steps in the RNN are correlated.

The products of nonlinearly transformed pre-activations $\phi_{i,\alpha}^{(a)}\equiv\phi(h_{i,\alpha}^{(a)})$
render the integrations in \prettyref{eq:gen_mom_gen_func} analytically
non-solvable. To find a suitable approximation, we insert auxiliary
variables in time ($a,b$) and in replica space ($\alpha,\beta$),
which account for the replica and time-step coupling. Introducing
these, the system decouples in the neuron indices $i$. We combine
RNN and DNN by defining the auxiliary variables

\begin{eqnarray}
C_{\alpha,\beta}^{(a,b)} & =M_{a,b}\bigg[ & \sigma_{a}^{2}+\frac{g_{a}^{2}}{n_{a-1}}\mathbb{1}_{a\ge1,b\ge1}\,\phi_{i,\alpha}^{(a-1)}\phi_{i,\beta}^{(b-1)}\nonumber \\
 &  & +\frac{g_{0}^{2}}{n_{\text{in}}}\mathbb{1}_{a\le A,b\le A}\,x_{i,\alpha}^{(a)}x_{i,\beta}^{(b)}\bigg]\label{eq:aux_field_appendix}
\end{eqnarray}
for $0\le a,b\le A+1$ with $M_{a,b}$ defined in \prettyref{eq:net_distinguishing_matrix},
$g_{a}=g$ for $1\le a\le A$ in RNN, and $n_{-1}\equiv n_{\text{in}}$.
The indicator functions $\mathbb{1}_{a\ge1,b\ge1}$ and $\mathbb{1}_{a\le A,b\le A}$
ensure that the respective terms vanish when they are not present,
e.g., the recurrent term $\phi_{i,\alpha}^{(a-1)}M_{a,b}\phi_{i,\beta}^{(b-1)}$
in the first step $a=b=0$. As above, there is an implicit sum over
the neuron indices $i$ on the right hand side.

We introduce these auxiliary variables by means of Dirac distributions
expressed as Fourier integrals
\begin{eqnarray}
\delta[\mathrm{\text{\prettyref{eq:aux_field_appendix}}}] & = & \prod_{\alpha,\beta=1}^{n_{D}}\,\prod_{a,b=0}^{A+1}\left\{ n_{a-1}\int_{i\mathbb{R}}\,\frac{d\tilde{C}_{\alpha,\beta}^{(a,b)}}{2\pi i}\right\} \nonumber \\
 &  & \times\exp\left(-\sum_{a,b=0}^{A+1}n_{a-1}\tilde{C}_{\alpha,\beta}^{(a,b)}\,(C_{\alpha,\beta}^{(a,b)}-\sigma_{a}^{2}\,M_{a,b}J_{\alpha,\beta})\right)\nonumber \\
 &  & \times\exp\left(\sum_{a,b=1}^{A+1}\tilde{C}_{\alpha,\beta}^{(a,b)}g_{a}^{2}\,\phi_{i,\alpha}^{(a-1)}M_{a,b}\phi_{i,\beta}^{(b-1)})\right)\nonumber \\
 &  & \times\exp\left(\sum_{a,b=0}^{A}n_{a-1}\tilde{C}_{\alpha,\beta}^{(a,b)}\frac{g_{0}^{2}}{n_{\text{in}}}x_{i,\alpha}^{(a)}M_{a,b}x_{i,\beta}^{(b)}\right),\label{eq:Dirac_aux_app}
\end{eqnarray}
where we inserted $J_{\alpha,\beta}=1$ for all $\alpha$ and $\beta$
to imply the summation over $\alpha,\beta$ that accounts for the
common biases across replicas. Used in the integrand of \prettyref{eq:gen_mom_gen_func},
this leads to
\begin{eqnarray}
p(\bm{Y}\,|\,\bm{X}) & = & \prod_{\alpha=1}^{n_{D}}\left\{ \int d\tilde{\bm{y}}_{\alpha}\int D\bm{h}_{\alpha}\int D\tilde{\bm{h}}_{\alpha}\right\} \prod_{\alpha,\beta=1}^{n_{D}}\left\{ \int D\tilde{C}_{\alpha,\beta}\int DC_{\alpha,\beta}\right\} \nonumber \\
 &  & \times\exp\left(\tilde{y}_{i,\alpha}y_{i,\alpha}+\frac{1}{2}\tilde{y}_{i,\alpha}C_{\alpha,\beta}^{(A+1,A+1)}\tilde{y}_{i,\beta}\right)\nonumber \\
 &  & \times\exp\left(\sum_{a=0}^{A}\tilde{h}_{i,\alpha}^{(a)}h_{i,\alpha}^{(a)}+\sum_{a,b=0}^{A}\frac{1}{2}\tilde{h}_{i,\alpha}^{(a)}C_{\alpha,\beta}^{(a,b)}\tilde{h}_{i,\beta}^{(b)}\right)\nonumber \\
 &  & \times\exp\left(-\sum_{a,b=0}^{A+1}n_{a-1}\tilde{C}_{\alpha,\beta}^{(a,b)}\,(C_{\alpha,\beta}^{(a,b)}-\sigma_{a}^{2}\,M_{a,b}J_{\alpha,\beta})\right)\nonumber \\
 &  & \times\exp\left(\sum_{a,b=1}^{A+1}\tilde{C}_{\alpha,\beta}^{(a,b)}g_{a}^{2}\phi_{i,\alpha}^{(a-1)}M_{a,b}\phi_{i,\beta}^{(b-1)})\right)\nonumber \\
 &  & \times\exp\left(\sum_{a,b=0}^{A}n_{a-1}\tilde{C}_{\alpha,\beta}^{(a,b)}\frac{g_{0}^{2}}{n_{\text{in}}}x_{i,\alpha}^{(a)}M_{a,b}x_{i,\beta}^{(b)}\right)\label{eq:p_Y_X}
\end{eqnarray}
with $DC_{\alpha,\beta}=\prod_{a,b=0}^{A+1}\,dC_{\alpha,\beta}^{(a,b)}$,
$D\tilde{C}_{\alpha,\beta}=\prod_{a,b=0}^{A+1}\frac{n_{a-1}d\tilde{C}_{\alpha,\beta}^{(a,b)}}{2\pi i}$.

We see in \prettyref{eq:p_Y_X} that there are no auxiliary variables
$C_{\alpha,\beta}^{(a,b)}$ that couple the output layer ($a=A+1$,
second line) with variables $\bm{h}_{\alpha}^{(a)},\tilde{\bm{h}}_{\alpha}^{(a)}$
in the rest of the network ($0\le a\le A$). This is a consequence
of the independence of the priors on the associated weights. We further
see in \prettyref{eq:p_Y_X} that no products of variables with different
neuron indices appear. The exponential thus factorizes into $n_{a}$
identical terms for each $a$. Rearranging the integrations, we obtain

\begin{eqnarray}
p(\bm{Y}\,|\,\bm{X})=\prod_{\alpha=1}^{n_{D}}\left\{ \int d\tilde{\bm{y}}_{\alpha}\right\}  & e^{\tilde{y}_{i,\alpha}y_{i,\alpha}} & \Big\langle e^{\frac{1}{2}\tilde{y}_{i,\alpha}C_{\alpha,\beta}^{(A+1,A+1)}\tilde{y}_{i,\beta}}\Big\rangle_{\tilde{C},C}\label{eq:p_y_prefinal}
\end{eqnarray}
where the expectation value is computed with respect to the action

\begin{equation}
\mathcal{S}_{\mathrm{aux}}(C,\tilde{C})=-n\sum_{a,b=0}^{A+1}\nu_{a-1}\tilde{C}_{\alpha,\beta}^{(a,b)}C_{\alpha,\beta}^{(a,b)}+n\,\mathcal{W}_{\text{aux}}(\tilde{C}\,|\,C)\label{eq: S_aux_app}
\end{equation}
of the auxiliary variables $\tilde{C}_{\alpha,\beta}^{(a,b)},C_{\alpha,\beta}^{(a,b)}$.
This action comprises the nontrivial part of the dynamics of the network
in the cumulant generating functional 
\begin{eqnarray}
\mathcal{W}_{\text{aux}}(\tilde{C}\,|\,C) & = & \frac{1}{n}\ln\left\langle e^{\sum_{a,b=1}^{A+1}\tilde{C}_{\alpha\beta}^{(a,b)}g_{a}^{2}\phi_{i,\alpha}^{(a-1)}M_{a,b}\phi_{i,\beta}^{(b-1)}}\right\rangle _{\{h_{i,\alpha}^{(a)}\}}\nonumber \\
 &  & +\sum_{a,b=0}^{A}\nu_{a-1}\tilde{C}_{\alpha,\beta}^{(a,b)}\frac{g_{0}^{2}}{n_{\text{in}}}x_{i,\alpha}^{(a)}M_{a,b}x_{i,\beta}^{(b)}\nonumber \\
 &  & +\sum_{a,b=0}^{A+1}\nu_{a-1}\tilde{C}_{\alpha,\beta}^{(a,a)}\sigma_{a}^{2}M_{a,b}J_{\alpha,\beta}\label{eq:W_rec_app}
\end{eqnarray}
where $\{h_{i,\alpha}^{(a)}\}$ describes the Gaussian statistics
of a single pre-activation $h_{i,\alpha}^{(a)}$ with covariance matrix
$\langle h_{i,\alpha}^{(a)}h_{i,\beta}^{(b)}\rangle=C_{\alpha,\beta}^{(a,b)}\delta_{i,j}$
across neurons $i,j$, time steps or layers $a,b$, and inputs $\alpha,\beta$.
Here $\nu_{a}=n_{a}/n$ denotes the relative layer sizes in the DNN.

To show that 
\begin{eqnarray}
\frac{1}{n}\ln\left\langle e^{\sum_{a,b=1}^{A+1}\tilde{C}_{\alpha\beta}^{(a,b)}g_{a}^{2}\phi_{i,\alpha}^{(a-1)}M_{a,b}\phi_{i,\beta}^{(b-1)}}\right\rangle _{\{h_{i,\alpha}^{(a)}\}} & = & O(1)\label{eq:scaling_W_aux_app}
\end{eqnarray}
and thus $\mathcal{W}_{\text{aux}}(\tilde{C}\,|\,C)=O(1)$, i.e.,
that $\mathcal{W}_{\text{aux}}(\tilde{C}\,|\,C)$ does not scale with
$n$, we consider RNN and DNN separately. For the RNN, the result
immediately follows because the neurons are uncorrelated, $\langle h_{i,\alpha}^{(a)}h_{i,\beta}^{(b)}\rangle=C_{\alpha,\beta}^{(a,b)}\delta_{i,j}$,
which factorizes the expectations and leads to a sum over $n$ identical
terms:
\begin{eqnarray*}
\frac{1}{n} & \ln & \Big\langle e^{\sum_{a,b=1}^{A+1}\tilde{C}_{\alpha\beta}^{(a,b)}g_{a}^{2}\phi_{i,\alpha}^{(a-1)}\phi_{i,\beta}^{(b-1)}}\Big\rangle_{\{h_{i,\alpha}^{(a)}\}}\\
 & = & \ln\Big\langle e^{\sum_{a,b=1}^{A+1}\tilde{C}_{\alpha\beta}^{(a,b)}g_{a}^{2}\phi_{\alpha}^{(a-1)}\phi_{\beta}^{(b-1)}}\Big\rangle_{\{h_{\alpha}^{(a)}\}}.
\end{eqnarray*}
For the DNN, one first notices that, by definition, $C_{\alpha,\beta}^{(a,b)}=0$
for $a\neq b$, so different layers decouple in \prettyref{eq:W_rec_app}.
Formally this can be seen by solving the integrals over the corresponding
variables $\tilde{C}_{\alpha,\beta}^{(a,b)}$ with $a\neq b$. This
factorization allows us to study each layer separately and decouple
the $n_{a}$ neurons:
\begin{eqnarray*}
\frac{1}{n} & \ln & \left\langle e^{\sum_{a=1}^{A+1}\tilde{C}_{\alpha\beta}^{(a,a)}g_{a}^{2}\phi_{i,\alpha}^{(a-1)}\phi_{i,\beta}^{(a-1)}}\right\rangle _{\{h_{i,\alpha}^{(a)}\}}\\
 & = & \sum_{a=1}^{A+1}\nu_{a-1}\ln\left\langle e^{\tilde{C}_{\alpha\beta}^{(a,a)}g_{a}^{2}\phi_{\alpha}^{(a-1)}\phi_{\beta}^{(a-1)}}\right\rangle _{\{h_{\alpha}^{(a)}\}}.
\end{eqnarray*}
Consequently, for both architectures \prettyref{eq:scaling_W_aux_app}
holds and $\mathcal{W}_{\text{aux}}(\tilde{C}\,|\,C)=O(1)$.

\subsubsection{Saddle-point approximation \label{app:Apendix_Saddle-point-approximation}}

The action $\mathcal{S}_{\mathrm{aux}}$ in the auxiliary fields scales
with the number of neurons $n$. In the limit $n\to\infty$, a saddle-point
approximation of the integrals over $\tilde{C}$ and $C$ appearing
in the expectation value in \prettyref{eq:p_y_prefinal} becomes exact.
The saddle points are determined by the stationary points of the action
$\mathcal{S}_{\mathrm{aux}}$ as $\frac{\partial}{\partial\tilde{C}_{\alpha,\beta}^{(a,b)}}\mathcal{S}_{\mathrm{aux}}(C,\tilde{C})\overset{!}{=}0$
and $\frac{\partial}{\partial C_{\alpha,\beta}^{(a,b)}}\mathcal{S}_{\mathrm{aux}}(C,\tilde{C})\overset{!}{=}0$,
leading to
\begin{eqnarray}
\overline{\tilde{C}}_{\alpha\beta}^{(a,b)} & = & 0,\\
\overline{C}_{\alpha\beta}^{(a,b)} & =M_{a,b}\bigg[ & \sigma_{a}^{2}+\frac{g_{0}^{2}}{n_{\text{in}}}\mathbb{1}_{a\le A,b\le A}x_{i,\alpha}^{(a)}x_{i,\beta}^{(b)}\nonumber \\
 &  & +g_{a}^{2}\mathbb{1}_{a\ge1,b\ge1}\langle\phi(h_{\alpha}^{(a-1)})\phi(h_{\beta}^{(b-1)})\rangle_{\{h_{\alpha}^{(a)}\}\sim\N(0,\overline{C})}\bigg]\label{eq:mf_with_replica_final_app}
\end{eqnarray}
with indices $a,b\in\{0,\dots,A+1\}$. The saddle point $\overline{\tilde{C}}_{\alpha\beta}^{(a,b)}=0$
is a self-consistent solution because $W_{\text{aux}}(0\,|\,C)\equiv0$,
which is in particular independent of $C$, so that $\partial W_{\text{aux}}(0\,|\,C)/\partial C_{\alpha,\beta}^{(a,b)}\equiv0$.

To evaluate the expectation value on the r.h.s.~of \prettyref{eq:mf_with_replica_final_app},
we only need the sub-tensors of $\overline{C}$ formed by the indices
that explicitly appear in the expectation due to the marginalization
property of the Gaussian. In particular, this means the saddle point
equations can be solved iteratively starting from $a=0$, which requires
the starting values $\overline{C}_{\alpha\beta}^{(0,a)}=\overline{C}_{\alpha\beta}^{(a,0)}=M_{a,0}\Big[\sigma_{0}^{2}+\frac{g_{0}^{2}}{n_{\text{in}}}\sum_{j=1}^{n_{\text{in}}}x_{j,\alpha}^{(a)}x_{j,\beta}^{(0)}\Big]$
for the recursion.

After the saddle-point approximation, the conditional probability
\prettyref{eq:p_y_prefinal} simplifies to the factorized Gaussian
\begin{eqnarray*}
p(\bm{Y}\,|\,\bm{X}) & = & \prod_{i=1}^{n_{A+1}}p(\bm{y}_{i}\,|\,\bm{X})\,,\\
p(\bm{y}_{i}\,|\,\bm{X}) & = & \N(0,\overline{C}^{(A+1,A+1)})\,,
\end{eqnarray*}
with covariance matrix $\langle y_{i,\alpha}y_{i,\beta}\rangle=\overline{C}{}_{\alpha,\beta}^{(A+1,A+1)}$
across inputs $\alpha,\beta$ that is determined recursively by \prettyref{eq:mf_with_replica_final_app},
starting from the input covariance $\overline{C}_{\alpha\beta}^{(0,0)}=\sigma_{0}^{2}+\frac{g_{0}^{2}}{n_{\text{in}}}\sum_{j=1}^{n_{\text{in}}}x_{j,\alpha}^{(0)}x_{j,\beta}^{(0)}$.
The diagonal elements $\overline{C}{}_{\alpha,\beta}^{(A+1,A+1)}$
thus only depends on the equal-time overlaps $\sum_{i=1}^{n_{\text{in}}}x_{i,\alpha}^{(a)}x_{i,\beta}^{(a)}$
of the inputs with $0\le a\le A$.

\subsection{Finite-size instability of RNNs\label{app:RNN_instability}}

\begin{figure}
\begin{centering}
\includegraphics{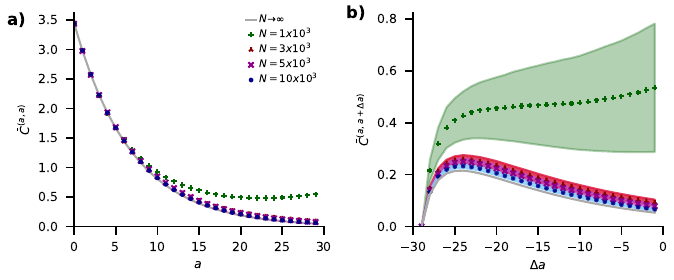}
\par\end{centering}

\caption{\textbf{Mean-field theory of the RNN compared to simulation. a }Average
variance in mean-field theory $\overline{C}{}^{(a,a)}$ (solid gray
curve) and estimate $\frac{1}{n}\sum_{i}h_{i}^{(a)}h_{i}^{(a)}$ from
simulation, averaged over $100$ realizations of networks with different
widths (symbols, see legend). \textbf{b }Average cross-covariance
in mean-field-theory $\overline{C}{}^{(a,a+\Delta a)}$ and estimate
$\frac{1}{n}\sum_{i}h_{i}^{(a)}h_{i}^{(a+\Delta a)}$ from simulation,
averaged over $100$ network realizations (mean shown as symbols,
same symbol code as in panel a; standard error of the mean shown as
tube), as a function of the temporal distance $\Delta a$ to the hidden
layer $a=30$. Mean-field theory (gray curve). Other parameters: $g^{2}=1.73$,
$\sigma^{2}=0$, layer widths $n_{a}\in\{1,3,5,10\}\cdot10^{3}$,
$A=30$ hidden layers, ReLU activation $\phi(x)=\max(0,x)$ and Gaussian
input $\bm{x}\protect\overset{\text{i.i.d.}}{\sim}\mathcal{N}(1,1)$
with $\text{dim}(\text{\ensuremath{\bm{x}}})=10^{5}$.\label{fig:finite_size_effect}}
\end{figure}

In the main text Figure 1, the mean-field theory is compared to network
simulations with hidden layer width $n=2000$ and a ReLU nonlinearity
for fixed hyperparameters $g^{2}=1.6,\,\sigma^{2}=0.2$. Although
this appears to be quite wide already, for the RNN the statistics
of the activity in individual networks strongly varies across realizations
of weights. The frequency of deviating realizations increases as one
approaches $g^{2}\to2$, the instability threshold above which $\overline{C}^{(a,a)}$
diverges with growing $a$ for the ReLU nonlinearity. The instability
threshold can be obtained from the MFT solution for a single replicon
and $a=b$, \prettyref{eq:MFT_relu_activity}: The theory predicts
that $g^{2}>2$ will lead to exponential increase of the activity,
while $g^{2}<2$ results in finite (but possibly very strong) activity.
Beyond this threshold, trajectories of individual neurons diverge
towards $\pm\infty$ over time. At finite width and $g^{2}<2$, the
activity is typically stable. But for $g^{2}$ sufficiently close
to $2$, the closeness of the instability point is visible in the
system. This is observable as a spread of individual neurons' trajectories,
each hovering about a non-zero set point. This observation corresponds
to a static contribution (independent of $\Delta a$) to the time-lagged
correlation function, as shown in \prettyref{fig:finite_size_effect}b.
The reason for this instability to only occur in the RNN is the coherent
interplay of the activity with the connectivity across time: Since
the connectivity is identical across all time steps, fluctuations
of the activity can be amplified coherently across multiple time steps.
Likewise, deviations of the variances $C^{(a,a)}$ are observable
in this case (\prettyref{fig:finite_size_effect}a). The effect is
suppressed as the network size increases; the mean-field theory then
becomes accurate also for values of $g^{2}$ close to $2$ (\prettyref{fig:finite_size_effect}a,b).

\subsection{Mean-field theory for error function nonlinearity\label{app:erf_appendix}}

\begin{figure}
\begin{centering}
\includegraphics{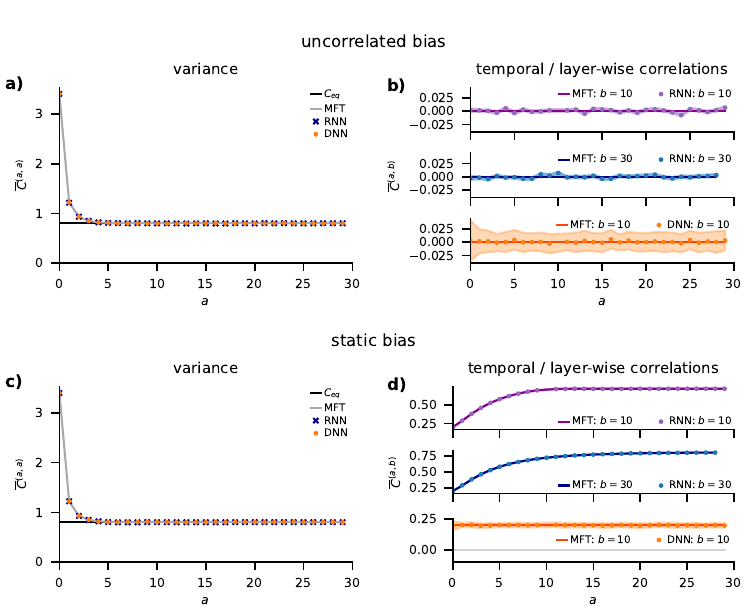}
\par\end{centering}
\caption{\textbf{Mean-field theory for DNN and RNN with a single input. a)
}Average variance in mean-field theory $\overline{C}{}^{(a,a)}$ (\prettyref{eq:MFT};
solid gray curve) and estimate $\frac{1}{n_{a}}\sum_{i}h_{i}^{(a)}h_{i}^{(a)}$
from simulation, averaged over $100$ realizations of networks, for
biases that are uncorrelated across time/layers (blue crosses RNN;
orange dots DNN). \textbf{b) }Cross-covariance $\overline{C}{}^{(a,b)}$
as a function of the hidden layer index $a$ for fixed $b\in\{10,30\}$
and uncorrelated biases. RNN: Mean-field theory (solid dark blue and
dark magenta). Mean (blue / purple dots) and standard error of the
mean (light blue / light purple tube) of $\frac{1}{n_{a}}\sum_{i}h_{i}^{(a)}h_{i}^{(b)}$
estimated from simulation of $100$ network realizations. DNN: Mean
(orange dots) and standard error of the mean of $\frac{1}{n_{a}}\sum_{i}h_{i}^{(a)}h_{i}^{(b)}$
estimated from simulation of $100$ network realizations. Other parameters
$g_{0}^{2}=g^{2}=1.6$, $\sigma^{2}=0.2$, finite layer width $n_{a}=2000$,
$A=30$ hidden layers, activation $\phi(x)=\mathrm{erf}(\sqrt{\pi}x/2)$
and Gaussian inputs $\bm{x}\protect\overset{\text{i.i.d.}}{\sim}\mathcal{N}(1,1)$
with $n_{\text{in}}=10^{5}$. \textbf{c)} Same as a) but for biases
that are static across time/layers. \textbf{d)} Same as b) but for
the static bias case.\label{fig:Mean-field-theory-for-DNN-RNN-erf}}
\end{figure}

While we calculated and discussed the temporal structure of the mean-field
kernel for the ReLU activation function in the main text, we here
focus on the odd activation function $\phi(x)=\mathrm{erf}(\sqrt{\pi}x/2)$
(the scaling ensures $\phi^{\prime}(0)=1$) and the single input case.
For the auto-correlation (\prettyref{fig:Mean-field-theory-for-DNN-RNN-erf}
panel a for uncorrelated bias and c for static bias), we see good
agreement with the theory, similar to $\phi=\mathrm{ReLU}$ case in
\prettyref{fig:Mean-field-theory-for-DNN-RNN} in the main text. The
temporal or layer-wise correlations are shown in panels b and d. As
discussed in \prettyref{sec:Kernel}, neither temporal nor layer-wise
correlations can arise in the uncorrelated bias case: For the DNN
this is clear for any activation function due to the independently
drawn weights, but also in the RNN case correlations vanish because
$\mathrm{erf}$ is an odd function. However, if temporal correlations
are induced via a static bias, as in panel d, these can be strengthened
by the weight sharing. In DNNs on the other hand, all correlations
can be accounted to the static bias applied in each layer. Note that
the fluctuations are smaller for $\phi=\mathrm{erf}$ than for $\phi=\mathrm{ReLU}$
(compare \prettyref{fig:Mean-field-theory-for-DNN-RNN} and \prettyref{fig:Mean-field-theory-for-DNN-RNN-erf}).

\subsection{Details about numerical experiments}

For all experiments, we used NumPy \cite{Harris20_357} and SciPy
\cite{Virtanen20_261} which are both released under a BSD-3-Clause
License. Computations were performed on a CPU cluster. More precisely,
the requirements for the experiments are:
\begin{itemize}
\item Figure 1 (main): 1h on a single core laptop.
\item Figure 2 (main): 50h on a single node with 24 cores of a CPU cluster
for each of panel a and c, and 2h on a single node with 24 cores of
the CPU cluster for each of panel b and d.
\item Figure 3 (main): 1h on a single core laptop.
\item Figure 4 (appendix): 1.5h on a single core laptop.
\item Figure 5 (appendix): 1h on a single core laptop.
\end{itemize}
The code used to produce the figures is stored in a Zenodo archive
with the DOI 10.5281/zenodo.5747219.

To solve the mean field theory for a given activation function $\phi(x)$
one needs to calculate the expectation values in \prettyref{eq:MFT},
or more general in \prettyref{eq:MFT_replicated}. Here we choose
the ReLU activation $\phi(x)=\mathrm{max}(0,x)$ as shown in \cite{Cho09}:
\begin{eqnarray}
\langle\phi(x)\phi(y)\rangle_{x,\,y\sim\N(0,C)} & = & \frac{1}{2\pi}\nu(\sin\theta+(\pi-\theta)\cos\theta)\,,\label{eq:ReLU_sol}
\end{eqnarray}
where
\begin{eqnarray*}
\nu & = & \sqrt{C_{xx}C_{yy}}\,,\\
\theta & = & \cos^{-1}\bigg(\frac{C_{xy}}{\nu}\bigg)\,.
\end{eqnarray*}
Inserting this into the MFT equations for multiple replicas \eqref{eq:MFT_replicated}
results in
\begin{eqnarray}
\overline{C}_{\alpha\beta}^{(a,b)} & =M_{a,b}\bigg[ & \sigma_{a}^{2}+\mathbb{1}_{a\ge1,b\ge1}\,\frac{g_{a}^{2}}{2\pi}\nu_{\alpha\beta}^{(a,b)}(\sin\theta_{\alpha\beta}^{(a,b)}+(\pi-\theta_{\alpha\beta}^{(a,b)})\cos\theta_{\alpha\beta}^{(a,b)})\nonumber \\
 &  & +\mathbb{1}_{a\le A,b\le A}\,\frac{g_{0}^{2}}{n_{\text{in}}}\bm{x}_{\alpha}^{(a)\T}\bm{x}_{\beta}^{(b)}\bigg]\,,\label{eq:ReLU_sol_MFT_replicated}
\end{eqnarray}
where
\begin{eqnarray*}
\nu_{\alpha\beta}^{(a,b)} & = & \sqrt{\overline{C}_{\alpha\alpha}^{(a-1,a-1)}\overline{C}_{\beta\beta}^{(b-1,b-1)}}\,,\\
\theta_{\alpha\beta}^{(a,b)} & = & \cos^{-1}\Bigg(\frac{\overline{C}_{\alpha\beta}^{(a-1,b-1)}}{\nu_{\alpha\beta}^{(a,b)}}\Bigg)\,.
\end{eqnarray*}
The special case $a=b$, $\alpha=\beta$, and vanishing external input
yields
\begin{eqnarray}
\overline{C}^{(a,a)} & = & \sigma_{a}^{2}+\frac{g_{a}^{2}}{2}\overline{C}^{(a-1,a-1)}\,.\label{eq:MFT_relu_activity}
\end{eqnarray}
For time or layer independent $g_{a}\equiv g$ and $\sigma_{a}\equiv\sigma$,
the activity thus increases exponentially in time or over layers for
$g^{2}>2$ and converges towards an equilibrium value $C^{(\infty)}=\frac{\sigma^{2}}{1-g^{2}/2}$
for $g^{2}<2$.

For the results based on $\phi(x)=\mathrm{erf}(\sqrt{\pi}x/2)$ shown
in \prettyref{app:erf_appendix}, we used \cite{Williams98_1203,vanMeegen21_043077}
\begin{eqnarray*}
\langle\phi(x)\phi(y)\rangle_{x,\,y\sim\N(0,C)} & = & \frac{2}{\pi}\,\arcsin\bigg(\frac{\frac{\pi}{2}C_{xy}}{\sqrt{1+\frac{\pi}{2}C_{xx}}\sqrt{1+\frac{\pi}{2}C_{yy}}}\bigg).
\end{eqnarray*}
Importantly, the r.h.s.~vanishes for uncorrelated inputs with $C_{xy}=0$.

\subsection{Next-to-leading-order corrections\label{app:Next-to-leading-order-correction}}

Here, we consider the case of a single input in the initial layer.
In this case, the action for the auxiliary variables $\mathcal{S}_{\mathrm{aux}}(C,\tilde{C})$
is given by \prettyref{eq:RNN_decoupled_action} together with $\mathcal{W}_{\text{aux}}^{\DNN}(\tilde{C}\,|\,C)$
from \prettyref{eq:W_rec_DNN} for DNNs and $\mathcal{W}_{\text{aux}}^{\RNN}(\tilde{C}\,|\,C)$
from \prettyref{eq:W_rec_RNN} for RNNs. To compute finite-size corrections,
we need so compute the Hessian of $\mathcal{S}_{\mathrm{aux}}(C,\tilde{C})$,
which is
\begin{eqnarray}
\mathcal{S}_{\mathrm{aux}}^{(a,b),(c,d)} & = & \left(\begin{array}{cc}
\mathcal{S}_{\mathrm{aux},11}^{(a,b),(c,d)} & \mathcal{S}_{\mathrm{aux},12}^{(a,b),(c,d)}\\
\mathcal{S}_{\mathrm{aux},21}^{(a,b),(c,d)} & \mathcal{S}_{\mathrm{aux},22}^{(a,b),(c,d)}
\end{array}\right)\nonumber \\
 & = & \left(\begin{array}{cc}
0 & \frac{\partial^{2}\mathcal{S}_{\mathrm{aux}}}{\partial C^{(a,b)}\,\partial\tilde{C}^{(c,d)}}\\
\frac{\partial^{2}\mathcal{S}_{\mathrm{aux}}}{\partial\tilde{C}^{(a,b)}\,\partial C^{(c,d)}} & \frac{\partial^{2}\mathcal{S}_{\mathrm{aux}}}{\partial\tilde{C}^{(a,b)}\,\partial\tilde{C}^{(c,d)}}
\end{array}\right),\label{eq:inv_prop}
\end{eqnarray}
where we used that due to the normalization of $\N(0,C)$ to one,
which is in particular independent of $C$, we have $\partial\mathcal{W}_{\text{aux}}/\partial C^{(a,b)}=0$
for $\tilde{C}=0$, so also $\partial^{2}\mathcal{S}_{\mathrm{aux}}/\partial C^{(a,b)}\,\partial C^{(c,d)}=0$.
Note that we here evaluate the fluctuations around the saddle-point
where $\overline{\tilde{C}}^{(a,b)}=0$ and $\overline{C}^{(a,b)}$
is given by \prettyref{eq:MFT}; to simplify the notation we drop
the overline throughout this subsection. To proceed, we separate the
derivation for DNN and RNN due to the difference between $\mathcal{W}_{\text{aux}}^{\DNN}(\tilde{C}\,|\,C)$
and $\mathcal{W}_{\text{aux}}^{\RNN}(\tilde{C}\,|\,C)$.

\subsubsection{DNN}

First, we focus on the DNN. The off-diagonal elements of the Hessian
are
\begin{eqnarray*}
\frac{\partial^{2}\mathcal{S}_{\mathrm{aux}}}{\partial C^{(a,b)}\,\partial\tilde{C}^{(c,d)}} & = & -n\,\nu_{a-1}\,\delta_{(a,b),(c,d)}+n\,\delta_{c,d}\,\nu_{c-1}g_{c}^{2}\,\frac{\partial\,\langle\phi^{(c-1)}\phi^{(c-1)}\rangle}{\partial C^{(a,b)}},
\end{eqnarray*}
where $\delta_{(a,b),(c,d)}=1$ if $a=c$ and $b=d$ and $\delta_{(a,b),(c,d)}=0$
otherwise. The derivative in the second term is the linear response
of the expectation value $\langle\phi^{(c-1)}\phi^{(c-1)}\rangle$
with regard to changes of $C^{(a,b)}$. Since the former only depends
on the statistics of $h^{(c-1)}$ and thus on $C^{(c-1,c-1)}$, we
obtain
\begin{eqnarray*}
\frac{\partial\,\langle\phi^{(c-1)}\phi^{(c-1)}\rangle}{\partial C^{(a,b)}} & = & \mathbb{1}_{0\le a\le A}\,\delta_{a,b}\,\delta_{a,c-1}\,\frac{\partial\,\langle\phi^{(a)}\phi^{(a)}\rangle}{\partial C^{(a,a)}}.
\end{eqnarray*}
Thus, we arrive at
\begin{eqnarray}
\frac{\partial^{2}\mathcal{S}_{\mathrm{aux}}}{\partial C^{(a,b)}\,\partial\tilde{C}^{(c,d)}} & = & -n\,\nu_{a-1}\,\delta_{(a,b),(c,d)}\nonumber \\
 &  & +n\,\mathbb{1}_{0\le a\le A}\,\delta_{a,b}\,\delta_{c,d}\,\delta_{a,c-1}\,\nu_{c-1}\,F_{c-1},\label{eq:S_off_12}\\
F_{c-1} & := & g_{c}^{2}\,\frac{\partial\,\langle\phi^{(c-1)}\phi^{(c-1)}\rangle}{\partial C^{(c-1,c-1)}}.\nonumber 
\end{eqnarray}
Analogously, the other off-diagonal element is
\begin{eqnarray}
\frac{\partial^{2}\mathcal{S}_{\mathrm{aux}}}{\partial\tilde{C}^{(a,b)}\,\partial C^{(c,d)}} & = & -n\,\nu_{a-1}\,\delta_{(a,b),(c,d)}\nonumber \\
 &  & +n\,\mathbb{1}_{0\le c\le A}\,\delta_{a,b}\,\delta_{c,d}\,\delta_{c,a-1}\text{\, \ensuremath{\nu_{a-1}}}\,F_{a-1}.\label{eq:S_off_21}
\end{eqnarray}
Finally, the diagonal element is
\begin{eqnarray*}
\frac{\partial^{2}\mathcal{S}_{\mathrm{aux}}}{\partial\tilde{C}^{(a,b)}\,\partial\tilde{C}^{(c,d)}} & = & \mathbb{1}_{a\ge1}\delta_{a,b}\,\delta_{a,c}\,\delta_{c,d}\,n\,\nu_{a-1}\,G_{a-1}\\
G_{a-1} & := & g_{a}^{4}\,\langle\phi^{(a-1)}\phi^{(a-1)}\phi^{(a-1)}\phi^{(a-1)}\rangle\\
 &  & -g_{a}^{4}\,\langle\phi^{(a-1)}\phi^{(a-1)}\rangle^{2}.
\end{eqnarray*}
With the Hessian at hand, we proceed to the propagator, i.e., the
inverse of the Hessian $-\sum_{k}\sum_{c,d}\mathcal{S}_{\mathrm{aux},ik}^{(a,b),(c,d)}\,\Delta_{kj}^{(c,d),(e,f)}=\delta_{(a,b),(e,f)}\delta_{i,j}$.

Due to the structure of the Hessian \eqref{eq:inv_prop}, the propagator
has the structure
\begin{eqnarray*}
\Delta & = & \left(\begin{array}{cc}
\Delta_{11} & \Delta_{12}\\
\Delta_{21} & 0
\end{array}\right).
\end{eqnarray*}
We first consider the off-diagonal elements $\Delta_{12}$, which
are time-reversed functions of one another $\Delta_{12}^{(a,b),(c,d)}=\Delta_{21}^{(c,d),(a,b)}$,
and can be determined from $-\mathcal{S}_{\mathrm{aux},21}\Delta_{12}=\mathbf{1}$
or, more explicitly,
\begin{eqnarray*}
-\sum_{c,d}\,\frac{\partial^{2}\mathcal{S}_{\mathrm{aux}}}{\partial\tilde{C}^{(a,b)}\,\partial C^{(c,d)}}\,\Delta_{12}^{(c,d),(e,f)} & = & \delta_{(a,b),(e,f)}.
\end{eqnarray*}
Using the explicit expression \eqref{eq:S_off_21} we obtain the Dyson
equation
\begin{eqnarray*}
\nu_{a-1}\,n\,\Delta_{12}^{(a,b),(e,f)} & - & n\,\mathbb{1}_{1\le a\le A+1}\,\delta_{a,b}\,\nu_{a-1}\,F_{a-1}\,\Delta_{12}^{(a-1,a-1),(e,f)}\\
 & = & \delta_{(a,b),(e,f)}.
\end{eqnarray*}
From here, we implicitly assume $1\le a\le A+1$ and drop the indicator
$\mathbb{1}_{1\le a\le A+1}$. Evaluating this equation for $a\neq b$
we get
\begin{eqnarray}
\Delta_{12}^{(a,b),(e,f)} & = & (\nu_{a-1}\,n)^{-1}\,\delta_{(a,b),(e,f)}\quad\text{for }a\neq b.\label{eq:initial_cond}
\end{eqnarray}
For $a=b$ one obtains an iterative equation
\begin{eqnarray*}
\nu_{a-1}\,n\,\Delta_{12}^{(a,a),(e,f)}-n\,\nu_{a-1}\,F_{a-1}\,\Delta_{12}^{(a-1,a-1),(e,f)} & = & \delta_{(a,a),(e,f)}.
\end{eqnarray*}
Due to linearity, a valid solution for $e\neq f$ is $\Delta_{12}\equiv0$.
For $e=f$ we obtain the single-index iteration
\begin{eqnarray}
\Delta_{12}^{(a,a),(e,e)} & = & (\nu_{a-1}\,n)^{-1}\delta_{(a,a),(e,e)}+F_{a-1}\,\Delta_{12}^{(a-1,a-1),(e,e)},\label{eq:iteration_single_index}
\end{eqnarray}
with the initial condition
\begin{eqnarray*}
\Delta_{12}^{(a,a),(a,a)} & = & (\nu_{a-1}\,n)^{-1}
\end{eqnarray*}
and for $a>e$ obeying the iteration
\begin{eqnarray*}
\Delta_{12}^{(a,a),(e,e)} & = & F_{a-1}\,\Delta_{12}^{(a-1,a-1),(e,e)}
\end{eqnarray*}
which has the solution
\begin{eqnarray}
\Delta_{12}^{(a,a),(e,e)} & = & \mathbb{1}_{a\ge e}n^{-1}\,\nu_{e-1}^{-1}\,\prod_{k=e}^{a-1}F_{k}\label{eq:iteration_response}
\end{eqnarray}
with $\prod_{k=e}^{e-1}F_{k}\equiv1$. Finally, we need to determine
$\Delta_{11}$ which obeys
\begin{eqnarray*}
-\sum_{c,d}\,\mathcal{S}_{\mathrm{aux},21}^{(a,b),(c,d)}\,\Delta_{11}^{(c,d),(e,f)}-\sum_{c,d}\,\mathcal{S}_{\mathrm{aux},22}^{(a,b),(c,d)}\,\Delta_{21}^{(c,d),(e,f)} & = & 0,\\
-\sum_{c,d}\,\mathcal{S}_{\mathrm{aux},11}^{(a,b),(c,d)}\,\Delta_{11}^{(c,d),(e,f)}-\sum_{c,d}\,\mathcal{S}_{\mathrm{aux},12}^{(a,b),(c,d)}\,\Delta_{21}^{(c,d),(e,f)} & = & \delta_{(a,b),(e,f)}.
\end{eqnarray*}
Since $\mathcal{S}_{\mathrm{aux},11}\equiv0$, the second equation
is fulfilled by the solution of $\Delta_{21}$ alone. The first equation
yields an additional condition for $\Delta_{11}$, which we obtain
by using $-\mathcal{S}_{\mathrm{aux},21}=\Delta_{12}^{-1}$. Thus,
multiplying from left with $\Delta_{12}$ yields
\begin{eqnarray}
\Delta_{11} & = & \Delta_{12}\,\mathcal{S}_{\mathrm{aux},22}\,\Delta_{21}.\label{eq:cov_from_prop}
\end{eqnarray}
Written explicitly, this results in
\begin{eqnarray*}
\Delta_{11}^{(a,b),(c,d)} & = & \sum_{(a^{\prime},b^{\prime}),(c^{\prime},d^{\prime})}\,\Delta_{12}^{(a,b),(a^{\prime},b^{\prime})}\,\mathcal{S}_{\mathrm{aux},22}^{(a^{\prime},b^{\prime}),(c^{\prime},d^{\prime})}\,\Delta_{21}^{(c^{\prime},d^{\prime}),(c,d)}\\
 & = & \sum_{a^{\prime}}\Delta_{12}^{(a,b),(a^{\prime},a^{\prime})}\,\Big[n\,\nu_{a^{\prime}-1}\,G_{a^{\prime}-1}\Big]\,\Delta_{21}^{(a^{\prime},a^{\prime}),(c,d)}.
\end{eqnarray*}
Only entries $\Delta_{12}$ with equal indices in the second pair
appear which fulfill the iteration \prettyref{eq:iteration_single_index}.
In addition, we have that $\Delta_{12}^{(a\neq b),(a^{\prime},a^{\prime})}\equiv0$
due to \prettyref{eq:initial_cond}. Thus, we arrive at
\begin{eqnarray}
\Delta_{11}^{(a,b),(c,d)} & = & n^{-1}\,\delta_{a,b}\delta_{c,d}\sum_{a'=1}^{\min(a,d)}\left\{ \prod_{k=a'}^{a-1}F_{k}\right\} \nu_{a'-1}^{-1}\,G_{a'-1}\left\{ \prod_{l=a'}^{d-1}F_{l}\right\} \label{eq:Delta_11_DNN}
\end{eqnarray}
For $a=b=c=d$, this simplifies to
\begin{eqnarray*}
\Delta_{11}^{(a,a),(a,a)} & = & n^{-1}\sum_{a'=1}^{a}\left\{ \prod_{k=a'}^{a-1}F_{k}^{2}\right\} \nu_{a'-1}^{-1}\,G_{a'-1}.
\end{eqnarray*}
To determine the solution, we need to evaluate $F_{a}$ and $G_{a}$.

For the ReLU activation function, we get from \prettyref{eq:MFT_relu_activity}
the simple relation $\langle\phi^{(a)}\phi^{(a)}\rangle=\frac{1}{2}C^{(a,a)}$
and thus 
\begin{eqnarray}
F_{a-1} & = & \frac{1}{2}g_{a}^{2}.\label{eq:F_ReLU}
\end{eqnarray}
For $G_{a}$ we also need $\langle\phi^{(a)}\phi^{(a)}\phi^{(a)}\phi^{(a)}\rangle=\frac{3}{2}C^{(a,a)}C^{(a,a)}$,
which follows from Wick's theorem, to obtain
\begin{eqnarray}
G_{a-1} & = & \frac{5}{4}g_{a}^{4}\,C^{(a-1,a-1)}C^{(a-1,a-1)}.\label{eq:G_ReLU}
\end{eqnarray}

\subsubsection{RNN}

The only difference between $\mathcal{W}_{\text{aux}}^{\DNN}$ and
$\mathcal{W}_{\text{aux}}^{\RNN}$ is that the latter does not decompose
into a sum, so the different time-points are not independent. The
elements of the Hessian \eqref{eq:inv_prop} therefore take the form
\begin{eqnarray*}
\frac{\partial^{2}\mathcal{S}_{\mathrm{aux}}}{\partial C^{(a,b)}\,\partial\tilde{C}^{(c,d)}} & = & -n\,\delta_{(a,b),(c,d)}+n\,g^{2}\,\frac{\partial\,\langle\phi^{(c-1)}\phi^{(d-1)}\rangle}{\partial C^{(a,b)}}.
\end{eqnarray*}
The derivative in the second term is the linear response of the expectation
value $\langle\phi^{(c-1)}\phi^{(d-1)}\rangle$ with regard to changes
of $C^{(a,b)}$. Since the former only depends on the joint statistics
of $h^{(c-1)},h^{(d-1)}$ and thus on $C^{(c-1,c-1)}$, $C^{(c-1,d-1)}$,
and $C^{(d-1,d-1)}$ the matrix elements of $\partial^{2}\mathcal{S}_{\mathrm{aux}}/\partial C^{(a,b)}\,\partial\tilde{C}^{(c,d)}$
are only non-zero in a $2\times2$ block where at most two different
indices appear in total. Analogously, the second term in
\begin{eqnarray*}
\frac{\partial^{2}\mathcal{S}_{\mathrm{aux}}}{\partial\tilde{C}^{(a,b)}\,\partial C^{(c,d)}} & = & -n\,\delta_{(a,b),(c,d)}+n\,g^{2}\,\frac{\partial\,\langle\phi^{(a-1)}\phi^{(b-1)}\rangle}{\partial C^{(c,d)}}
\end{eqnarray*}
depends only on $C^{(a-1,a-1)}$, $C^{(a-1,b-1)}$, and $C^{(b-1,b-1)}$
which leads again to a $2\times2$ block structure. The diagonal element
is

\begin{eqnarray*}
\frac{\partial^{2}\mathcal{S}_{\mathrm{aux}}}{\partial\tilde{C}^{(a,b)}\,\partial\tilde{C}^{(c,d)}} & = & n\,g^{4}\,\langle\phi^{(a-1)}\phi^{(b-1)}\phi^{(c-1)}\phi^{(d-1)}\rangle\\
 &  & -n\,g^{4}\,\langle\phi^{(a-1)}\phi^{(b-1)}\rangle\langle\phi^{(c-1)}\phi^{(d-1)}\rangle\\
 & =: & n\,G_{a-1,b-1,c-1,d-1},
\end{eqnarray*}
which does not vanish for any combination of parameters.

The resulting propagator again obeys
\begin{eqnarray*}
-\sum_{c,d}\,\frac{\partial^{2}\mathcal{S}_{\mathrm{aux}}}{\partial\tilde{C}^{(a,b)}\,\partial C^{(c,d)}}\,\Delta_{12}^{(c,d),(e,f)} & = & \delta_{(a,b),(e,f)},
\end{eqnarray*}
which explicitly reads
\begin{eqnarray*}
n\,\Delta_{12}^{(a,b),(e,f)} & = & \delta_{(a,b),(e,f)}+n\,g^{2}\sum_{c,d}\,\frac{\partial\,\langle\phi^{(a-1)}\phi^{(b-1)}\rangle}{\partial C^{(c,d)}}\,\Delta_{12}^{(c,d),(e,f)}.
\end{eqnarray*}
We restrict ourselves to $a=b=c=d$, i.e., $\Delta_{11}^{(a,a),(a,a)}$,
for which it is by \prettyref{eq:cov_from_prop} sufficient to determine
$\Delta_{12}^{(a,a),(e,f)}$. For $a=b$, we obtain
\begin{eqnarray*}
n\,\Delta_{12}^{(a,a),(e,f)} & = & \delta_{(a,a),(e,f)}+n\,\mathbb{1}_{1\le a\le A+1}\,F_{a-1}\,\Delta_{12}^{(a-1,a-1),(e,f)}
\end{eqnarray*}
because $g^{2}\,\partial\,\langle\phi^{(a-1)}\phi^{(a-1)}\rangle/\partial C^{(d,c)}=g^{2}\,\delta_{a-1,c}\,\delta_{a-1,d}\,F_{a-1}$.
Thus, for $e=f$, we recovered \prettyref{eq:iteration_single_index}
which is solved by \prettyref{eq:iteration_response}. For $e\neq f$,
$\Delta_{12}^{(a,a),(e,f)}\equiv0$ due to the vanishing inhomogeneity.

The same expression \eqref{eq:cov_from_prop} holds as in the case
of the DNN,
\begin{eqnarray*}
\Delta_{11}^{(a,b),(c,d)} & = & \sum_{(a^{\prime},b^{\prime}),(c^{\prime},d^{\prime})}\,\Delta_{12}^{(a,b),(a^{\prime},b^{\prime})}\,\mathcal{S}_{\mathrm{aux},22}^{(a^{\prime},b^{\prime}),(c^{\prime},d^{\prime})}\,\Delta_{21}^{(c^{\prime},d^{\prime}),(c,d)}.
\end{eqnarray*}
For $\Delta_{11}^{(a,a),(a,a)}$ we obtain
\begin{eqnarray}
\Delta_{11}^{(a,a),(a,a)} & = & n^{-1}\,\sum_{a^{\prime},c^{\prime}=1}^{a}\,\big\{\prod_{k=a'}^{a-1}F_{k}\big\}\,G_{a^{\prime}-1,a^{\prime}-1,c^{\prime}-1,c^{\prime}-1}\,\big\{\prod_{l=c'}^{a-1}F_{l}\big\}\label{eq:Delta_11_RNN}
\end{eqnarray}
because $\Delta_{12}^{(a,a),(e,f)}=0$.

For the ReLU activation function, we need \cite{Cho09}
\begin{eqnarray}
\langle\phi^{(a)}\phi^{(a)}\phi^{(b)}\phi^{(b)}\rangle & = & \frac{3}{2\pi}\nu_{a,b}^{2}\sin\theta_{a,b}\cos\theta_{a,b}\nonumber \\
 &  & +\frac{1}{2\pi}\nu_{a,b}^{2}(\pi-\theta_{a,b})(1+2\cos^{2}\theta_{a,b})\label{eq:quartic_relu_average}
\end{eqnarray}
with $\nu_{a,b}=\sqrt{C^{(a,a)}C^{(b,b)}}$ and $\theta_{a,b}=\cos^{-1}\bigg(\frac{C^{(a,b)}}{\nu_{a,b}}\bigg)$
as in \prettyref{eq:ReLU_sol}. With this, we can determine
\begin{eqnarray*}
G_{a-1,a-1,c-1,c-1} & = & g^{4}\,\langle\phi^{(a-1)}\phi^{(a-1)}\phi^{(c-1)}\phi^{(c-1)}\rangle\\
 &  & -\frac{1}{4}g^{4}\,C^{(a-1,a-1)}C^{(c-1,c-1)}.
\end{eqnarray*}

\providecommand{\newblock}{}

\end{document}